\documentclass[10pt]{iopart}

\usepackage{graphicx}
\newcommand{\homega}{\hat{\bOmega}}
\newcommand{\psiarg}{(\mathbf{r},\homega,E)}

 \usepackage{iopams}
\usepackage{multirow}
\begin{document}
\title[Deterministic Neutronics for Fusion]{A novel discontinuous-Galerkin deterministic neutronics model for Fusion applications: development and benchmarking}
\author{Timo Bogaarts$^{1, *}$ and Felix Warmer$^{1,2}$}
\address{$^1$ Eindhoven University of Technology, PO Box 513, 5600 MB, Eindhoven, Netherlands}
\address{$^2$ Max-Planck-Institut für Plasmaphysik, Teilinstitut Greifswald, Wendelsteinstrasse 1, Greifswald D-17491, Germany}
\address{$^*$ Corresponding author: t.j.bogaarts@tue.nl}
\date{August 2024}
\begin{abstract}
Neutron interactions in a fusion power plant play a pivotal role in determining critical design parameters such as coil-plasma distance and breeding blanket composition. Fast predictive neutronic capabilities are therefore crucial for an efficient design process. For this purpose, we have developed a new deterministic neutronics method, capable of quickly and accurately assessing the neutron response of a fusion reactor, even in three-dimensional geometry. It uses a novel combination of arbitrary-order discontinuous Galerkin spatial discretization, discrete-ordinates angular and multigroup energy discretizations, arbitrary-order anisotropic scattering, and matrix-free iterative solvers, allowing for fast and accurate solutions. One, two, and three-dimensional models are implemented. Cross sections can be obtained from standard databases or from Monte-Carlo simulations. Benchmarks and literature tests were performed, concluding with a successful blanket simulation. 
\end{abstract}
\submitto{\NF}

\maketitle

\ioptwocol 
\section{Introduction}
The neutron response of a fusion power plant determines key performance characteristics such as tritium breeding ratio, coil lifetime, and maintenance requirements. Therefore, fusion reactor design must include neutron modelling of critical systems such as the blanket and coils, even in the early design stages. These stages include both systems-code driven design \cite{lion2021general, kembleton2022eu} and parametric reactor studies \cite{franza2015implementation, moreno2024parastell}.

Currently, such early-stage fusion reactor design mostly uses either conventional stochastic Monte-Carlo methods \cite{valentine2023advancing, palermo2024challenges} or reduced models \cite{franza2015implementation, kovari2016process}. 

Although potentially very accurate, Monte-Carlo methods often need $10^8-10^9$ samples to converge, requiring thousands of CPU hours on supercomputers to simulate the full geometry. Furthermore, simulating the curved and complicated Computer-Aided-Design (CAD) geometry can also be challenging, requiring external programs for meshing and simulating the faceted geometry or approximations to the actual geometry using constructive solid geometry. The reduced models are on the other side of the extreme: although very fast, they are incapable of resolving the full three-dimensional geometry required for a more detailed design (or in the case of a stellarator, a design at all) beyond the zero or one-dimensional level (as produced by the systems codes in which they are used).

In fusion reactor design, and especially early-stage design and optimization, it is desirable to be able to quickly assess neutronic viability of a design in order to have a fast design cycle. Therefore, although both types of models have advantages and disadvantages, they can be sub-optimal for design: either very accurate, three-dimensional, and too expensive, or fast, and too inaccurate for or incapable of three-dimensional modelling. Hence, for design and particularly early-stage design or optimization, a model that fills the space between these extremes could be useful.

Such a model should be usable in a systems code or in a standalone manner (useful for parametric design or possibly more detailed design work), require minimal user-input, and be able to model the three-dimensional geometry quickly and accurately enough for design purposes. In this work, a model that has the potential to satisfy these requirements is derived, implemented, and tested. In contrast to stochastic Monte-Carlo methods, it directly solves the six-dimensional neutron transport equation in a deterministic fashion. Although in principle less accurate than converged stochastic methods due to the used six-dimensional discretization, valid results can be obtained in substantially less computational time. In addition, the full distribution function is immediately obtained, which can be of interest in the design process (e.g., by adding extra shielding in high neutron fluence or penetration regions). Finally, the method is based on an unstructured mesh, such that analyses can be performed on a wide range of geometries: from highly-detailed, single components, to full reactor designs (simplified, parametric, or detailed). This unstructured mesh also allows results to be relatively straightforwardly coupled to other design-relevant engineering codes if so desired. 

Deterministic models have been extensively investigated before \cite{Lewis1984, evans2010denovo, wareing2001discontinuous, goldberg2022validating, moller2011minaret, fournier2013discontinuous, hall2017hp}, with also some applications to fusion \cite{royston2018application, youssef2007comparing}. However, to the best of our knowledge, the combination of the high-order discontinuous Galerkin unstructured mesh method, the unified 1D/2D/3D model, the efficient spatial transport sweep, and the matrix-free iterative methods used for within-group solution, is novel. The three-dimensional unstructured mesh is required for retaining the flexibility of application, while the other characteristics allow the scheme to be as fast as possible, which is benificial for the design process.

The preceding discussion especially holds for stellarators. A reduced model is by definition incapable of modelling the three-dimensional, inhomogeneous neutron response in the blanket. Furthermore, stochastic Monte-Carlo methods are particularly challenging due to the inherently curved blanket geometry \cite{haussler2017verification,haussler2018neutronics} and high sampling requirements due to the three-dimensional solution \cite{lyytinen2024proof}, even for simplified models of the blanket. Therefore, the model should be particularly useful for stellarator design. 

In section \ref{sec:numerical_models}, the numerical model is derived from the neutron transport equation, using a standard and well-known discrete-ordinates, multigroup, velocity domain discretization and a high-order, discontinuous Galerkin discretization. Then, in section \ref{sec:verification}, the resulting model and its convergence properties are verified in three tests: an analytical solution, a test with isotropic scattering, and a test with anisotropic scattering. In section \ref{sec:breedingblanket} a successful comparison with a Monte-Carlo code in a fusion-relevant application to a breeding blanket is then shown, in both one- and three-dimensional geometry. Finally, a conclusion and discussion are given in section \ref{sec:conclusion}.

\section{Numerical Models}\label{sec:numerical_models}
The fixed-source neutron transport equation is a six-dimensional Boltzmann transport equation and can be written as 
\begin{equation}\label{eq:neutron_transport_eq}
    \hat{\bOmega}\cdot \nabla\psi +\sigma^t  \, \psi= q^e + q^s(\psi),
\end{equation}
where $\homega$ denotes neutron streaming direction,  $E$ neutron energy, $\psi = \psi \psiarg$ the angular flux, $\sigma^t$ the total macroscopic cross section, $q^e$ the external neutron source (a source inside the computational domain independent of the angular neutron flux, e.g. a fusion plasma), and $q^s$ the source arising from neutron to neutron scattering (scattering source), which depends (linearly) on the angular flux itself.
First the assumptions made are explained. Then, the phase-space discretization and the scattering source computation are discussed. The used schemes are well-known, but will be explained shortly for completeness and for providing the context for the novel discontinuous-Galerkin spatial discretization that is presented afterwards.
\subsection{Assumptions}\label{sec:assumptions}
For the derivation of the algorithm, several assumptions are made:
\begin{itemize}
    \item Neutrons in fusion applications do not gain energy from scattering events \cite{royston2018application}. This simplifies the energy discretization scheme significantly by solving from high to low energy in one iteration.
    \item Elements used in the discontinuous-Galerkin discretization have straight edges and are convex. This is necessary for the efficient transport sweep algorithm of section \ref{sec:dg}.
    \item Each element is assumed to consist of one material (a sum of discrete nuclides and their densities). This will allow taking the cross sections out of finite-element integrals.
\end{itemize}
Note that as long as the physical system considered obeys the first assumption, this does not impact the accuracy of the method (e.g. fusion systems obey this assumption, fission systems do not and should thus not be simulated using this method). This first assumption does break down in the thermal energy range \cite{davison1958neutron}, as neutrons can gain energy from the thermal motion of the colliding atom. As the considered fusion systems operate at maximum on the order of $10^3$ K, this thermal effect only becomes appreciable at the eV energy range. Furthermore, this effect is included within the energy group itself: if the neutron upscatters but remains in the energy group, this is included in the scattering cross sections. Only neutrons upscattering from one energy group to a different energy group (group-to-group upscatter) are neglected. In the realistic breeding blanket benchmark in section \ref{sec:breedingblanket}, this assumption does not have a significant effect on the results.

The effect of the second and third assumptions can be mitigated by refining the mesh. The effect of the first assumption can be eliminated by using outer iterations over the (thermal) energy groups, but this is comes at a significant computational cost and is not used in the present work.

\subsection{Phase-space discretization}\label{sec:phase_space}
The energy and angular phase-space dependencies are discretized using the standard multi-group, discrete-ordinates approach \cite{Lewis1984}. In short, the energy is discretized by solving for the solution in a set of energy groups $E_g$ and the angular space is discretized by solving only on a number of discrete angles (ordinates).

An energy group $E_g$ is defined by its lower and upper bound. The $G$ energy groups are ordered such that $E_1$ has the highest energy and $E_G$ has the lowest energy. The neutron transport equation \eref{eq:neutron_transport_eq} for a specific energy group $E_g$ is then
\begin{equation}\label{eq:single_eg}
    \hat{\bOmega}\cdot \nabla\psi_g +\sigma^t_g  \, \psi_g= q^e_g + q^S_g (\{\psi_{\tilde{g}}\})
\end{equation}
where the subscript $g$ denotes the value at energy group $E_g$ (i.e., $\psi_g = \psi(\mathbf{r}, \hat{\bOmega}, E_g)$, and $q^S_g$ is the scattering source. This scattering source depends in principle on the angular flux for other energy groups, such that the equations for all energy groups $E_g$ are coupled to each other.

However, as mentioned in section \ref{sec:assumptions}, it is assumed that neutrons undergoing a scattering reaction do not gain energy. Therefore, the highest energy group is independent from all other energy groups and can be solved immediately. Then, the second highest energy group only depends on the highest energy group and can be solved when the highest energy group is known. This continues until the lowest energy group is solved. In effect, coupling between the energy-discretized equations is lower-triangular and can be solved with forward substitution \cite{evans2010denovo}. Thus, each energy group only contains scattering source contributions from higher energy groups and the group itself (note that higher energy groups mean lower indices: $E_1$ is the highest energy group). This motivates the modification of equation \eref{eq:single_eg} to
\begin{equation}\label{eq:energy_discretized}
        \hat{\bOmega}\cdot \nabla\psi_g +\sigma^t_g  \, \psi_g= q^e_g + q^D_g (\{\psi_{1}, \, \dots, \psi_{g-1}\}) + q^I_g (\psi_{g}),
\end{equation}
where $q^D_g$ is the downscatter source that depends only on the angular flux in higher energy groups, and $q^I_g$ is the inscatter source which depends only on the angular flux in energy group $E_g$ itself. Note that as $q^D_g$ does not depend on the flux in energy group $E_g$, it is effectively an external source during the solution of energy group $E_g$.

Then, the discrete ordinates method requires that this equation holds for a set of discrete angles $\hat{\bOmega}_k$:
\begin{equation}\label{eq:spatial_pde2}
    \hat{\bOmega}_k\cdot \nabla\psi_{kg} +\sigma^t_{g}  \, \psi_{kg}= q^e_{kg} + q^D_{kg} +  q^I_{kg}(\psi_g)\ \forall \bOmega_k
\end{equation}
where the subscript $k$ denotes that quantity in the direction $\bOmega_k$. The subscript $kg$ denotes thus a quantity in both the direction $\bOmega_k$ and in the energy group $E_g$: $f_{kg} = f(\bOmega_k, E_g)$. As the inscatter source $q^I_{kg}$ depends on the angular flux in general (and not only in that particular direction), this term couples the discrete directions. The sets of equations for all directions can be solved by iteratively solving the equation for all angles while updating the inscatter source \cite{Lewis1984}. However, this scattering iteration is rather inefficient, as for scattering-dominated problems it can take many iterations to converge. Therefore, modern matrix-free iterative solvers have also been implemented in a similar fashion as \cite{evans2010denovo}. They will be discussed in more detail in section \ref{sec:matrixfree}.

Before discussing the spatial discretization of equation \eref{eq:spatial_pde2}, the scattering source computation is explained.

\subsection{Scattering Source}\label{sec:scattering_source}
The differential scattering cross section $\sigma^s(\mathbf{r}, \tilde{\bOmega}\cdot \hat{\bOmega}, E_{\tilde{g}} \to E_g)$ is a function of the change of angle $\tilde{\bOmega}\cdot \hat{\bOmega}$. Although this dependence could allow us to directly scatter flux from one direction $\hat{\bOmega}_k$ to all others, this requires the storage of the solution of all angles in all energy groups, which can be prohibitive. Instead, as will be shown below, if the angular dependence is expanded in Legendre polynomials, only a relatively few number of spherical harmonic moments 
\begin{equation}\label{eq:moment_def}
 \psi^g_{lm} =\int \psi^g(\hat{\bOmega}) Y_{lm}(\hat{\bOmega})d\hat{\bOmega},
\end{equation}
with $Y_{lm}$ the $lm$ spherical harmonic, are required to calculate the scattering source, drastically reducing memory requirements.

Thus, the differential scattering cross section is expanded with Legendre polynomials $P_l$ as \cite{Lewis1984, evans2010denovo,  goldberg2022validating, boyd2019multigroup}
\begin{equation}
    \sigma^s(\mathbf{r},  \tilde{\bOmega}\cdot \homega, E_{\tilde{g}}\to E_g) \approx \sum_{l=0}^L P_l(\tilde{\bOmega}\cdot \homega)\sigma^l_{g\tilde{g}}(\mathbf{r}),
\end{equation}
where $E_{\tilde{g}}$ is the incoming energy group, $E_{{g}}$ is the outgoing energy group, $L$ is the maximum order considered ($L=0$ is isotropic scattering and $L\to \infty$ is exact), and $\sigma^l_{g\tilde{g}}$ is the differential scattering cross section expansion coefficient. $\sigma^0_{g\tilde{g}}$ is the scalar cross section for scattering from group $E_{\tilde{g}}$ to group $E_g$ while higher orders are corrections from isotropic scattering. Using the spherical harmonic addition theorem and the fact that neutrons only scatter from higher groups to lower groups (downscatter) or from an energy group to the same energy group (inscatter), the scattering source can then be rewritten as (see \ref{app:deriv_scat_source} for a derivation)
\begin{eqnarray}
    q^s(\mathbf{r}, E_g, \hat{\bOmega}) &= q^D_g +q^I_g \nonumber \\
    &=\sum_{\tilde{g} = 1}^{g-1} \sum_{l=0}^L \sum_{m=-l}^l  \sigma^l_{g\tilde{g}} Y^*_{lm}(\hat{\bOmega}) \phi^{\tilde{g}}_{lm}\nonumber \\
    &\phantom{\sum_{\tilde{g} =1}^{g-1} }+\sum_{l=0}^L \sum_{m=-l}^l  \sigma^l_{gg} Y^*_{lm}(\hat{\bOmega}) \phi^{g}_{lm}\label{eq:scat_split}
\end{eqnarray}
where $q_s^D$ is the downscatter contribution, $q_s^I$ the inscatter contribution, and $\phi^{\tilde{g}}_{lm}$ the $lm$ moment of the angular flux as (as in equation \eref{eq:moment_def}):
\begin{equation}\label{eq:philm_calculation}
    \phi^{\tilde{g}}_{lm} = \int \psi^{\tilde{g}}(\mathbf{r}, \tilde{\bOmega}) Y_{lm}(\tilde{\bOmega}) d\tilde{\bOmega},
\end{equation}
Since a number of discrete angles are used instead of a continuous description, the integral has to be approximated using a quadrature:
\begin{equation}\label{eq:psilm} \phi^{\tilde{g}}_{lm} \approx \sum_{k=1}^Q w_k \psi^{\tilde{g}}(\mathbf{r}, \tilde{\bOmega}_k) Y_{lm}(\tilde{\bOmega}_k). \end{equation}
 where $w_k$ is the weight of the discrete angle $\tilde{\bOmega}_k$ in the angular quadrature set with $Q$ angles. The spherical harmonic normalization used here is such that $Y_{00} \equiv 1$ and $\phi^{\tilde{g}}_{00}$ is the scalar flux. 

Now, write the $lm$ moments of the downscatter and inscatter contributions as 
\begin{eqnarray}
    q^D_{glm}(\mathbf{r}, E_g) = \sum_{\tilde{g} = 1}^{g-1}  \sigma^l_{g\tilde{g}} \phi^{\tilde{g}}_{lm}, \\ q^I_{glm}(\mathbf{r}, E_g) = \sigma^l_{gg} \phi^g_{lm} 
\end{eqnarray}
 such that the final form of the scattering source can be written as 
 \begin{equation}
     q^s_{kg}(\mathbf{r}) = \sum_{lm} Y^*_{lm}(\bOmega_k) \left[q^D_{glm} (\mathbf{r})+ q^I_{glm}(\mathbf{r})\right],
 \end{equation}
 where it is still a function of the spatial variable, which will be discretized in the next section.
\subsection{Spatial discretization}\label{sec:dg}
The equation to be spatially discretized is equation \eref{eq:spatial_pde2}, with the scattering and external source combined in a total source $q^T_{kg}$(as they will use the same spatial discretization):
\begin{equation}\label{eq:dg_disc_eq}
    \hat{\bOmega}_k\cdot \nabla\psi_{kg} +\sigma^t_{g}  \, \psi_{kg}= q^T_{kg}
\end{equation}
Note that this equation needs to be solved for all angles $\bOmega_k$ and energy groups $E_g$, such that a fast solution of this equation is crucial for the performance of the method. 

The spatial discretization is done using arbitrary order Discontinuous Galerkin methods \cite{hesthaven2007nodal}. Higher order elements can reduce both computational cost and memory requirements, while Discontinuous Galerkin methods allow for a fast solution by computing only one element at a time in a so-called sweep, as shown below. 

In contrast to standard finite-element methods, each element has its own basis functions. These basis functions $H_j$ have a finite value only inside the element they belong to. The angular flux and source can then be expanded as 
\begin{eqnarray}
    \psi(\mathbf{r}) &= \sum_j \psi_j H_j(\mathbf{r}), \\ q(\mathbf{r})    &= \sum_j q_j H_j(\mathbf{r})
\end{eqnarray}
where $\psi_j$ and $q_j$ are the expansion coefficients.

As in standard Galerkin methods, the set of test functions $\{H_i\}$ is set equal to the set of basis functions $\{H_j\}$, and equation \eref{eq:dg_disc_eq} is multiplied with such a test function $H_i$ and integrated over the domain:
\begin{eqnarray}
    \sum_j  \psi_j \int  H_i \hat{\bOmega} \cdot \nabla H_j &+ \sigma H_i H_j dV \nonumber \\
    &=  \sum_j q_j \int H_i H_j dV.\label{eq:dgfem}
\end{eqnarray}
Requiring this equation to hold for all $i$ yields the linear system ($q_j$ is known)
\begin{equation}\sum_j A_{ij} \psi_j = B_i\end{equation}
which could be solved. However, as the basis and test functions only have finite values inside the element they belong to, the integration over the whole domain and all basis functions reduces to an integral over the single element $M$ to which both $H_i$ and $H_j$ belong. Therefore, the are no matrix elements connecting different elements, and only single-element equations are solved. This does not allow flux to propagate and is therefore undesired. 

In order to remedy this, $\psi$ is not expanded just yet: 
\begin{equation}\label{eq:dgfem_non_exp}
    \int  H_i \hat{\bOmega} \cdot \nabla \psi + \sigma H_i H_j \psi dV =  \sum_j q_j \int H_i H_j dV.
\end{equation}
Now, the first term is partially integrated over the element $M$ to which $H_i$ belongs:
\begin{eqnarray}
    \int_M H_i \hat{\bOmega}\cdot \nabla \psi dV = &- \int_M \psi \hat{\bOmega} \cdot \nabla H_i \  dV  \nonumber\\& + \int_{\partial M} H_i \overline{\psi}\ \hat{\bOmega} \cdot \hat{\mathbf{n}} \ dA\label{eq:partial_int}
\end{eqnarray}
where $\overline{\psi}$ is the angular flux on the boundary $\partial M$ of the element $M$. Since the basis functions are not assumed to be continuous along element boundaries, there is an ambiguity on whether to use the angular flux of this element $M$ or the connected element $O$. This ambiguity can be exploited to obtain an advantageous solution scheme. In this case, the upwind numerical flux is used:
\begin{eqnarray}
    \overline{\psi}(\mathbf{r}) =  \cases{ \psi^M(\mathbf{r}) & $\hat{\bOmega}\cdot \hat{\mathbf{n}} > 0$\\
                    \psi^O(\mathbf{r}) & $\hat{\bOmega}\cdot \hat{\mathbf{n}} < 0$\\}    
\end{eqnarray}
The flux inside the element is used if the edge is an outflow boundary, and the flux inside the connected element is used if the edge is an inflow boundary. This numerical flux is consistent; if the discontinuous solution is replaced by the continuous solution, it reduces to simply $\psi$ along the surface.

As mentioned in section \ref{sec:assumptions}, only elements with straight edges are considered, such that $\hat{\bOmega}\cdot \hat{\mathbf{n}}$ is constant. Then, the total integral over all edges $e$ is: 
\begin{eqnarray}
    \int_{\partial M} H_i \overline{\psi}\ \hat{\bOmega} \cdot \hat{\mathbf{n}} \ dA &= \sum_{e^+} \hat{\bOmega}\cdot \hat{\mathbf{n}}_e \int_e H_i \psi^M dA \nonumber \\
    &+\sum_{e^-}\hat{\bOmega}\cdot \hat{\mathbf{n}}_e \int_e H_i \psi^O dA,
\end{eqnarray}
where $e^+$ denotes an outflow edge ($\hat{\bOmega} \cdot \hat{\mathbf{n}} > 0$) and $e^-$ an inflow edge ( $\hat{\bOmega} \cdot \hat{\mathbf{n}} <0$). Combining this with equation \eref{eq:partial_int} , replacing the first term in equation \eref{eq:dgfem_non_exp} with the result, and reintroducing the expansion coefficients, the linear system inside each element can be written as:
\begin{eqnarray}
    \fl \sum_j &\left[\int_M \sigma H_i H_j dV  - \int_M H_j \hat{\bOmega} \cdot \nabla H_i dV \right.  
\nonumber \\
&\left.+ \sum_{e^+} |\hat{\bOmega}\cdot \hat{\mathbf{n}}_e |\int_e H_i H_j dA\right]\psi_j \nonumber \\
    =& \phantom{+}\sum_{e^-} |\hat{\bOmega}\cdot \hat{\mathbf{n}}_e| \sum_l \psi^O_l   \int_e H_i \overline{H}_l dA \nonumber  \\ &+ \sum_j q_j \int  H_i H_j dV\label{eq:total_equation}    
\end{eqnarray}
where $\psi^O_l, \{\overline{H}_l\} $ are the expansion coefficients and basis functions from the connected elements $O$, and $\psi_j, \{H_i\},\{H_j\}$ are the expansion coefficients and basis functions from the base element $M$. Note that this equation should hold for all $H_i$ belonging to element $M$, and is therefore a matrix equation for each independent element. Furthermore, in each element, only the term containing $\psi^O_l$ depends on other elements, and, if it is known, element $M$ can be solved on its own.

\begin{figure}
    \centering
    \includegraphics[width=\linewidth]{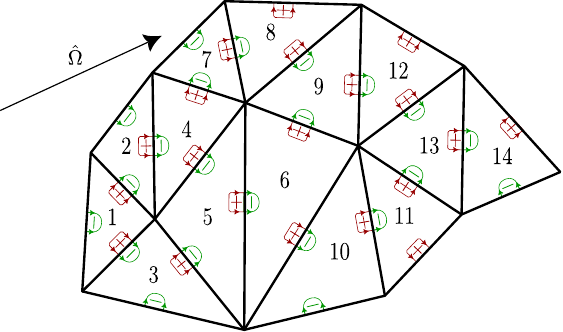}
    \caption{A numbered triangular mesh, together with a solution direction $\bOmega$. Boundaries corresponding to outflow boundaries (defined as $\hat{\bOmega} \cdot \hat{\mathbf{n}} > 0$) are indicated by a red plus, while inflow boundaries ($\hat{\bOmega} \cdot \hat{\mathbf{n}} < 0$) are indicated by a green minus. }
    \label{fig:graph_mesh}
\end{figure}

\begin{figure}
    \centering
    \includegraphics[width=\linewidth]{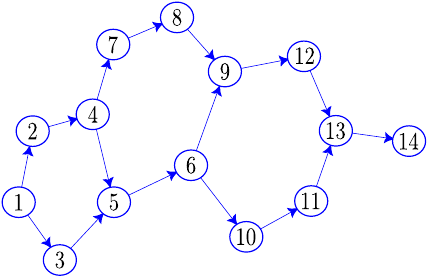}
    \caption{The directed acyclic graph obtained from the mesh in figure \ref{fig:graph_mesh} if one connects the mesh elements through their outflow and inflow boundaries. Topologically sorting can yield a solution order as $[1,2,3,4,5,6,7,8,9,10,11,12,13,14]$, although this is not unique. }
    \label{fig:graph_only}
\end{figure}

Now, consider figure \ref{fig:graph_mesh}, where a numbered triangular mesh is shown together with a direction $\hat{\bOmega}$. For each triangle the respective inflow and outflow boundaries are also shown. As element number 1 has no inflow boundaries connected to other elements, its right-hand-side does not depend on other elements (no $\psi^O_l$ term) and can be solved immediately. For element 2, the only inflow boundary is connected to element 1, such that it can be solved immediately after element 1 has been solved. Mapping the dependencies of elements to the solution of other elements gives the directed acyclic graph shown in figure \ref{fig:graph_only}. 

Topologically sorting this graph yields a solution order $\mathbf{C}_k$ (dependent on the direction $\bOmega_k$) in which elements can be solved one-by-one. In this simple case, it can be done by hand, but in larger computational meshes, this topological sorting can be performed using a depth-first search \cite{aho1974design}. 

Thus, the solution can be computed by sequentially solving $N$ small systems of size $p \times p$ ($N$ number of elements, $p$ number of unknowns per element) instead of solving a total system of $pN \times pN$, massively reducing solution times and memory usage. This 'sweep' algorithm is well-known \cite{evans2010denovo, joseph2005comparison, wareing2001discontinuous}, but the very general discontinuous-Galerkin formulation of equation \eref{eq:total_equation} is novel. 

\subsubsection{Scattering source computation}

Finally, the scattering source needs to be connected to the solution of the angular flux expansion coefficients. This is done by expanding the scattering source on the same basis:
\begin{eqnarray}
    q^s_{kg}(\mathbf{r})&= \sum_j q^s_{kgj}H_j(\mathbf{r}) \nonumber \\ 
                        &=\sum_j \left[\sum_{lm}Y^*_{lm}(\hat{\bOmega}_k)\left(q^D_{glmj} + q^I_{glmj}\right)\right]H_j(\mathbf{r})
\end{eqnarray}
with the coefficients $q^D_{glmj}, q^I_{glmj}$ being calculated as:
\begin{eqnarray}
\phi^{g}_{lmj} = \sum_{k} w_k \psi^{g}_{kj}Y_{lm}(\hat{\bOmega}_k)\label{eq:psilmj}\\
    q^D_{glmj} = \sum_{\tilde{g} = 1}^{g-1}  \sigma^l_{g\tilde{g}} \phi^{\tilde{g}}_{lmj}\label{eq:downscatter} \\ 
    q^I_{glmj} = \sigma^l_{gg} \phi^g_{lmj}\label{eq:inscatter}
\end{eqnarray}

\subsubsection{Parallel computation of elements}

As mentioned, equation \ref{eq:total_equation} is a matrix equation for each element. Due to the term containing $\psi^O_l$, they cannot be solved completely in parallel but have to be solved in the solution order $\mathbf{C}_k$. However, if we define the following matrices
\begin{eqnarray}
    \mathbf{G}_M &= G_{ij}^M = \int_M \sigma H_i H_j dV  - \int_M H_j \hat{\bOmega} \cdot \nabla H_i dV \nonumber \\
                                         &\phantom{\int_M \sigma H_i }\ \ \ \ \ \ \ \ \ \ \ \ \ \  +\sum_{e^+} |\hat{\bOmega}\cdot \hat{\mathbf{n}}_e |\int_e H_i H_j dA \label{eq:GM}\\
    \mathbf{Q}_M &= Q^M_i    =  \sum_j q_j \int  H_i H_j dV  \label{eq:QM}\\
    \mathbf{Q}_O &= Q^O_i    =\sum_{e^-} |\hat{\bOmega}\cdot \hat{\mathbf{n}}_e| \sum_l \psi^O_l   \int_e H_i \overline{H}_l dA \label{eq:QO}\\ 
    \bpsi^M &= \psi_j,
\end{eqnarray}
and write the total linear system \eref{eq:total_equation} as 
\begin{equation}
    \mathbf{G}_M\cdot  \bpsi_M = \mathbf{Q}_O + \mathbf{Q}_M,
\end{equation}
with the solution 
\begin{equation}
    \bpsi_M = \mathbf{G}_M^{-1} \cdot \left(\mathbf{O}_O + \mathbf{Q}_M\right),
\end{equation}
we note that only the relatively cheap matrix-vector product depends on other elements $O$. Thus, the relatively expensive matrix assembly and inversion $\mathbf{G}^{-1}_M$ can be done without information from other elements and therefore potentially in parallel.

\subsection{Total algorithm}\label{sec:total_algo}
\begin{figure*}
    \centering
    \includegraphics[width=\linewidth]{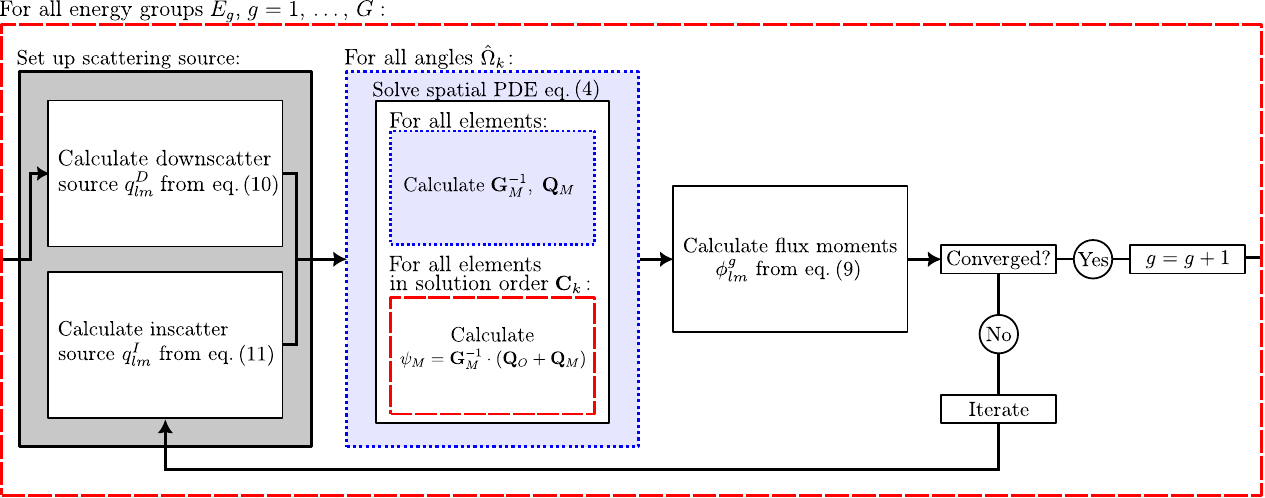}
    \caption{The total algorithm for solving the neutron transport equation \eref{eq:neutron_transport_eq} using the multigroup, discrete-ordinates, Discontinuous Galerkin, discretization. Red dashed boxes indicate sequential loops, while blue dotted boxes denote parallel loops. For each energy group $E_g$, the downscatter source is calculated, after which for all angles $\hat{\bOmega}_k$ (in parallel), the spatial PDE of equation \eref{eq:dg_disc_eq} is solved. This is done by (potentially in parallel) calculating for each element $M$ the independent matrices $\mathbf{G}^{-1}_M$, $\mathbf{Q}_M$, and then in sequential order calculating the matrix-vector products to obtain the solution inside each element $\psi_M$. Then, the flux moments $\phi_{lm}^g$ are calculated and convergence of these values are checked. If it is not converged, the inscatter source is calculated (either with the moments $\phi_{lm}^g$ or using the iterative solvers of section \ref{sec:matrixfree}) and the angle loop is restarted. If it is converged, the next energy group is calculated.}
    \label{fig:flowchart}
\end{figure*}

The total algorithm for general, straight-sided, convex elements is shown in figure \ref{fig:flowchart}. For each energy group, in sequential order, starting from the highest energy group $E_1$, first the downscattering source is computed. Then, in parallel, for each angle $\hat{\bOmega}$, the spatial PDE \ref{eq:dg_disc_eq} is solved using the sweep algorithm of section \ref{sec:dg}. Thus, for each element $M$, the matrices $\mathbf{G}_M^{-1}$, $\mathbf{Q}_M$ are calculated (potentially in parallel), after which the solution is obtained by a matrix-vector product in sequential order. Then, from the solution of the angular flux in all directions, the flux moments $\phi^g_{lm}$ are obtained from equation \eref{eq:psilmj}, with convergence checked. If not converged, the inscatter source is calculated from equation \eref{eq:inscatter} and equation \eref{eq:dg_disc_eq} is solved for all angles again (in the next section \ref{sec:matrixfree}, more advanced iterative solvers are explained, but the principle is similar). If it is converged, the algorithm proceeds with the next energy group $g+1$.

Computing the downscatter, inscatter and $\phi_{lm}$ moments can be done independently for each element. These steps thus scale very well with the number of processors up to the number of elements. Furthermore, they do not comprise a significant fraction of the total computational cost.

\subsection{Matrix free iterative solvers}\label{sec:matrixfree}
During an iteration, simply calculating the inscatter source with the solution of the moments $\psi^g_{lm}$ of the previous iteration, is although straightforward, not ideal. This iteration on the scattering source can converge quite slowly, needing upwards of 60 iterations for inscatter-dominated problem (e.g. 90\% of neutrons re-scattering in the same energy group). To improve on this, iterative solvers that calculate the inscatter source with a more informed choices have been used \cite{evans2010denovo, patton2002application}, and will be explained here as well. 

Note that $\psi_{kj}$, the angular flux solution values at each angle $\bOmega_k$ and in each elements solutions values $j$, are not necessarily the desired unknowns: the downscatter and inscatter source only depend on $\phi_{lmj}$, the $lm$ moments of the angular flux (see \eref{eq:psilmj}) in each elements solution values $j$. Furthermore, the desired quantities such as tritium breeding or neutron flux at the coil only depend on $\phi_{00j}$. Finally, $\phi_{lmj}$ contains much less values than $\psi_{kj}$: a high value of $L=5$ only has 36 moments compared to the several hundred discrete angles commonly used. As shown in detail in \ref{app:matrix_free}, the problem for one energy group can also be written as the linear system for $\phi_{lmj}$
\begin{eqnarray}\label{eq:a_op_def}
    \left(\mathbf{I} - \mathbf{M}\circ \mathbf{T}^{-1}\circ \mathbf{S}\right)[\phi_{lmj}] = \mathbf{M}\circ \mathbf{T}^{-1}[Q_{kj}] \equiv b_{lmj}
\end{eqnarray}
with $\mathbf{I}$ the identity operator, $\mathbf{M}$ the operator converting the discrete angle solution $\psi_{kj}$ to moments $\phi_{lmj}$, $\mathbf{T}$ the transport operator ($\hat{\bOmega}\cdot \nabla + \sigma$) in the discontinuous Galerkin formulation, $\mathbf{S}$ the operator converting moments $\phi_{lmj}$ to inscatter source $q^I_{kj}$ at the angle $\hat{\bOmega}_k$ and value $j$, and $Q_{kj}$ the external and downscatter source. The inverse $T^{-1}$ of the transport operator is precisely what is calculated during a sweep, and thus the action of this operator is relatively cheap to compute. The $b_{lmj}$ vector is obtained by simply sweeping for all angles using the external and downscatter source and computing the resulting $lm$ moments.

The left-hand side is a linear operator on the $\phi_{lmj}$ solution vector while the right-hand-side is a constant vector of values. Rewrite it as a linear equation as
\begin{eqnarray}
    \mathbf{A}\cdot \bphi = \mathbf{b}.
\end{eqnarray}
with $\mathbf{A} \equiv  \left(\mathbf{I} - \mathbf{M}\circ \mathbf{T}^{-1}\circ \mathbf{S}\right)$ the linear operator and $\mathbf{b} \equiv  b_{lmj}$ the right-hand-side.

Then, in principle, the matrix representation of the operator $\mathbf{A}$ could be constructed and inverted. However, to limit memory usage and computational time, matrix-free iterative solvers are used instead. They only require matrix-vector products of the type $\mathbf{A}\cdot \bphi$, or equivalently, the action of the operator $\mathbf{A}$ on $\bphi$, to iteratively solve the linear system, which, for large systems as considered here (note that although the number of elements is not too large, this has to be multiplied by the number of moments for the total system), can be significantly more efficient than direct solving \cite{saad2003iterative}. Thus, they do not require construction of the matrix representation of $\mathbf{A}$, only the definition of the operator $\mathbf{A}$ itself, limiting memory usage. Precisely this action of the operator $\mathbf{A}$ on an arbitrary given moment solution vector $\bphi$ can be computed quickly in the following steps (refer to equation \eref{eq:a_op_def}): 
\begin{itemize}
    \item Calculate scattering source using the given moments $\bphi$ ($\mathbf{S}$).
    \item Sweep over all angles using only the inscatter source as source ($\mathbf{T}^{-1}$)
    \item Calculate the moments $\overline{\bphi}$ resulting from this sweep ($\mathbf{M})$
    \item Return the vector $\bphi - \overline{\bphi}$
\end{itemize}

As an example of a matrix-free iterative solver, fixed point or Richardson iteration iteratively finds the solution by iterating as 
\begin{eqnarray}
    \bphi^{(k+1)} = \bphi^{(k)} + (\mathbf{b} - \mathbf{A}\bphi^{(k)}) 
\end{eqnarray}
where the superscript indicates the iteration number.
Using the definition for $\mathbf{A}$ and $\mathbf{b}$, this is equal to 
\begin{eqnarray}
    \bphi^{(k+1)} &= \mathbf{M}\circ \mathbf{T}^{-1}[Q_{gki}]  + \mathbf{M}\circ \mathbf{T}^{-1}\circ \mathbf{S}[\bphi^{(k)}]\nonumber \\
                  &= \mathbf{M}\circ \mathbf{T}^{-1}[Q_{gki} + \mathbf{S}[\bphi^{(k)}]].
\end{eqnarray}
This is exactly the scatter source iteration discussed in section \ref{sec:total_algo} and shown in figure \ref{fig:flowchart}: in each iteration, the inscatter source is computed with the solution of the previous iteration, swept over all angles, and converted to moment solution. However, as mentioned, this is not ideal and two other types of iterative solvers have been implemented: GMRES \cite{saad1986gmres} and BiCGSTAB \cite{van1992bi}.  These similarly update the $\bphi$ iteration vector using only applications of the operator $\mathbf{A}$ to the moments $\phi$.

Both GMRES and BiCGSTAB significantly reduce the number of sweeps required, often by more than a factor five, even without any preconditioning. GMRES requires storing all moment vectors from previous iterations and thus requires some overhead. However, the number of moments is much less than the number of angles and thus a total moment vector $\bphi$ requires much less memory than the total memory use from the sweep over the angles (i.e. a $\psi_{gkj}$ vector) and the overhead is therefore not as significant as it may seem. For BiCGSTAB, its memory requirement does not increase with the number of iterations, and although two applications of the operator $\mathbf{A}$ are required, the number of iterations is also reduced compared to GMRES, such that it is still competitive. 

Consequently, if memory requirements are a limit; scattering iteration is preferred. Otherwise; GMRES and BiCGSTAB are greatly preferred, with neither strongly outperforming the other.

\subsection{Simplification for simplexes}
Although the algorithm of the previous section can be used for general, straight sided, convex elements, significant simplifications are obtained under the assumption that the elements are (linear) simplexes. Then, as shown in \ref{app:simplices}, the single-element equation \eref{eq:total_equation} can be written as:
\begin{equation}
    \sum_j G^M_{ij} \cdot \psi_j^M = Q^O_{i} +  Q^M_{i}
\end{equation}
with now ($\boldsymbol{\mathcal{J}}_v$, $\mathcal{J}_v$, $\mathcal{J}_e$ are Jacobian (matrices) for transformations to base element coordinates)
\begin{eqnarray}
    G^M_{ij} &= \sigma \mathcal{J}_v \mathcal{A}_{ij} + \sum_{\alpha,\beta} \mathcal{J}_v \hat{\bOmega}_\beta  (\mathbf{J}_v^{-1})_{\alpha \beta}\mathcal{B}_{ij\alpha}\nonumber \\
    & \ \ \ \ \ \ \ \ \ \ \ \ \ \ + \sum_{e^+}\mathcal{J}_e |\hat{\bOmega}\cdot \hat{\mathbf{n}}_e|  \mathcal{C}_{ij}^e  \\     
    Q^O_i &=\sum_{e^-}\mathcal{J}_e  |\hat{\bOmega}\cdot \hat{\mathbf{n}}_e |\sum_j \mathcal{D}^{ev}_{ij}\psi^O_j  \\
    Q^M_i &=  \mathcal{J}_v\sum_j q_j \mathcal{A}_{ij}  
\end{eqnarray}
where the calligraphic matrices $\{\mathcal{A}_{ij}, \mathcal{B}_{ij\alpha}$, $\mathcal{C}^e_{ij}$, $\mathcal{D}^{ev}_{ij}\}$ do not depend on the element and can be retrieved from a central location. Thus, during a sweep, no integrals (i.e., the terms in equations \eref{eq:GM}, \eref{eq:QM}, \eref{eq:QO}) over elements or edges need to be calculated. The required matrices $\mathbf{G}_M, \mathbf{Q}_O, \mathbf{Q_M}$ can instead be obtained by simple matrix additions and multiplications. Crucially, this scheme does not depend on the order of the basis functions considered, such that even for high-order elements, no integrals have to be calculated using relatively expensive numerical techniques as numerical quadrature, considerably speeding up single-element calculations.

Currently, Lagrangian basis functions are used, defined on the Legendre-Gauss-Lobatto points to guarantee well-behavedness for higher order polynomials \cite{ern2004theory, hesthaven2007nodal}. This also allows for the implementation of boundary conditions by considering the values on the nodes on the edge of the elements, as required for the transport sweep. The calculation of the calligraphic matrices is automated and has been implemented for one-dimensional finite-elements, triangular finite elements, and tetrahedral finite elements, all with arbitrary element order. 
\subsection{Code description} 
The code is written in modern C++ and uses the Eigen \cite{eigenweb} library for matrix computations. The JSON file format can be used for configuration files while the HDF5 file format is used for data input and output. Python bindings are provided using the \texttt{nanobind} \cite{nanobind} binding library. 

The generic total algorithm of section \ref{sec:total_algo} does not require knowledge of the specific type of elements. Therefore, it has been implemented for all types of elements using templates. Each specific type of element merely needs to specify the computation of the $\mathbf{G}_M, \mathbf{Q}_O, \mathbf{Q_M}$ matrices and the creation of the domain. This makes it relatively easy for new types of elements to be implemented. 

At the moment, parallelization is performed using OpenMP. Both parallelization strategies (over the angles and over the elements, see figure \ref{fig:flowchart}) have been implemented, but only the parallelization over the angles has been used in this work. The precise scaling for realistic applications using either one or both strategies is the subject of future work. 

The basic scattering source iteration does not require an external solver. For the matrix-free iterative solvers, the GMRES and BiCGSTAB solvers of the Eigen library are used. 

Cross sections can be obtained from standard libraries such as FENDL \cite{schnabel2024fendl} or from Monte-Carlo runs \cite{boyd2019multigroup}.

\section{Verification of models}\label{sec:verification}
In order to verify the numerical method, several tests are performed, comparing the deterministic code with analytical solutions and Monte-Carlo simulations with the multigroup mode of the OpenMC code \cite{romano2015openmc}. In this multigroup mode, the equations sampled correspond exactly to the equations that are discretized in the deterministic model. Therefore, the simulations can, in principle, be brought arbitrarily close to each other such that convergence properties can be investigated. 

First, the spatial discretization and angular quadrature are verified by considering a simple 1D analytical solution. Then, two literature tests are performed to test isotropic and anisotropic scattering, with convergence assessed for both angular and spatial dimensions.


\subsection{Spatial and angular verification}\label{sec:analytical}
To verify the spatial and angular models, a slab geometry is used. The solution is investigated as a function of $x$ and the $y,z$ dimensions of the slab are chosen to be large enough to eliminate edge effects. The size of the slab in the $x$-direction is set to 1, while in the other dimensions it is set to 10. The one-dimensional neutron transport equation in direction $\hat{\bOmega}$ is
    \begin{eqnarray}\label{eq:single-angle}
    \Omega_x \frac{\partial \psi}{\partial x } = q - \sigma_t \psi ,
\end{eqnarray}
which has the solution
\begin{equation}\label{eq:single_angle_solution}     \psi =  \frac{q}{\sigma} \cases{1 - exp(-\sigma x / \Omega_x) & $\Omega_x> 0$\\ 
1 - exp(-\sigma (x - 1) / \bOmega_x) & $\Omega_x < 0$}
\end{equation}
where $q$ is the source, and $\sigma$ is the cross section. Both are set to 10 here. Furthermore, the analytical $\phi_{lm}$ moments can be obtained by integrating this solution using equation \eref{eq:philm_calculation}. The slab is discretized in a number of blocks, each of which contains 6 tetrahedrons. In the $yz$ direction this number is 11, while in the $x$-direction it is varied to probe the convergence properties. The quadrature sets used are all from the TN family \cite{thurgood1995tn}.

For the spatial verification, an angle with $\Omega_x = 0.24$ was chosen and the convergence was investigated for different finite element orders.
For the angular verification, the spatial discretization was set to 60 blocks in the $x$ direction to guarantee that the spatial discretization error is smaller than the angular quadrature error (in fact, the analytical solution \ref{eq:single_angle_solution} has been directly integrated with the numerical quadrature set as well, obtaining the same result). Then, the $L^2$ error of the numerical solution of the $\{\phi_{00}, \phi_{10}, \phi_{20}, \phi_{30}, \phi_{40}\}$ moments as a function of the number of angles was calculated.

\begin{figure*}
    \centering
    \includegraphics[width=\linewidth]{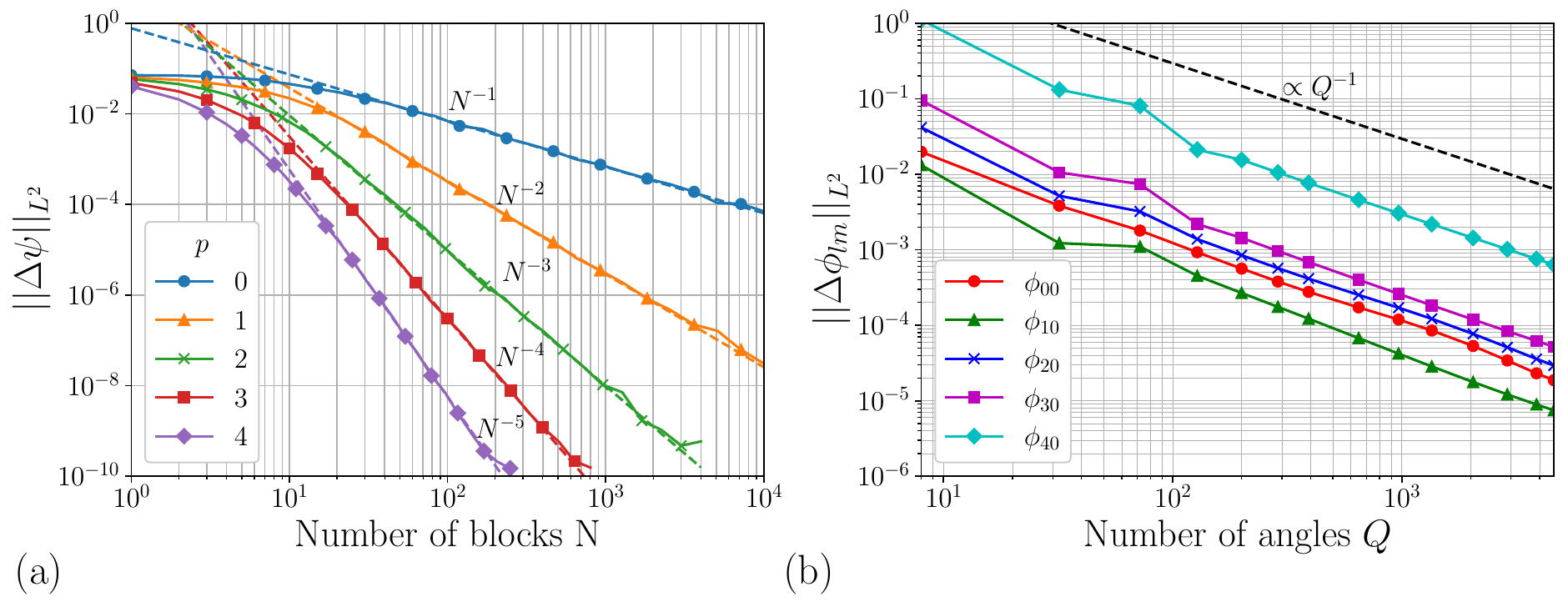}
    \caption{The $L^2$ error for the numerical solution of equation \eref{eq:single-angle} with the tetrahedral finite elements as a function of the number of finite element blocks at $\Omega_x = 0.24 $ (a) The order $p$ of the finite element is indicated in the plot. All orders show the optimal $N^{p+1}$ convergence rate. }
    \label{fig:4_analytical_conv}
\end{figure*}

Both spatial and angular convergence are shown in figure \ref{fig:4_analytical_conv}.  For the spatial convergence, all order show the optimal \cite{hesthaven2007nodal} $N^{p+1}$ convergence rate. For the triangular and one-dimensional finite elements, the same test was performed and also there the optimal convergence rates are attained. For the angular convergence, all $lm$ moments show the same $Q^{-1}$ convergence rate.

The results from this section show that the spatial elements are correctly implemented and that higher order elements can reproduce accurate results with much fewer elements than the conventional lower order elements. Furthermore, the angular integration has been correctly implemented, converging with a linear convergence rate in this particular problem. Finally, note that in general, accuracy is a function of both spatial discretization and quadrature set, with both possibly limiting total accuracy.


\subsection{Isotropic scattering test}
\begin{figure}[ht]
    \centering
    \includegraphics[width=\linewidth]{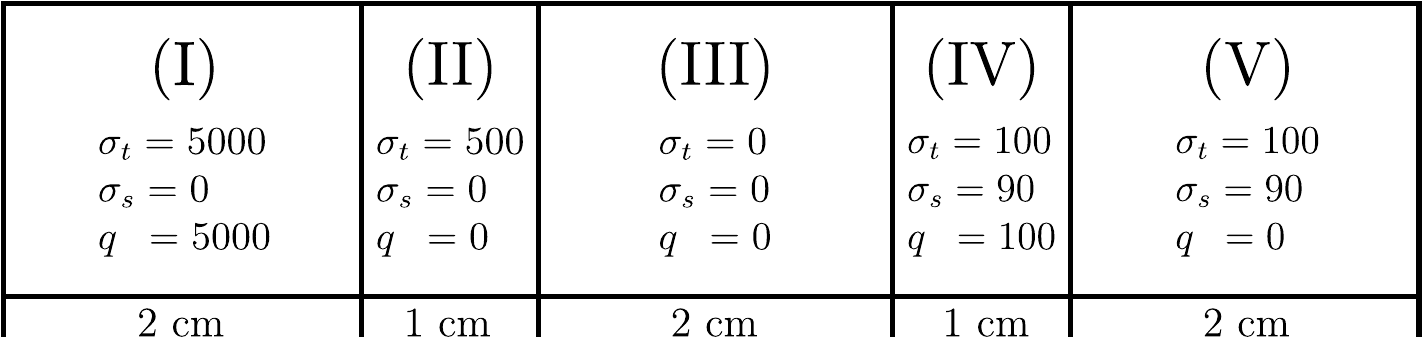}
    \caption{The domain setup for the isotropic scattering test. $\sigma_t$ is the total cross section in $m^{-1}$, $\sigma_s$ the isotropic scattering cross section in $m^{-1}$, and $q$ the isotropic source in $m^{-3}$. Data from Ref. \cite{reed1971new}.}
    \label{fig:reed_setup}
\end{figure}

The isotropic scattering test performed here was first described in Ref. \cite{reed1971new}. It is a single energy-group, one-dimensional, multi-region test using a total cross section $\sigma_t$, a scattering cross section $\sigma_s$ (i.e. $\sigma^l_{g\tilde{g}} = \sigma^0_{00} = \sigma_s$), and an isotropic source $q$. The specific values and the domain geometry are shown in figure \ref{fig:reed_setup}. The comparison is done with the multigroup mode of OpenMC, since this will solve the exact same equations (in a different manner), and thus convergence can be assessed. Although it is a one-dimensional test, the codes are run in three-dimensional geometries, with the size of the other dimensions being 1 meter to eliminate edge effects.

\begin{figure}[ht]
    \centering
    \includegraphics[width=\linewidth]{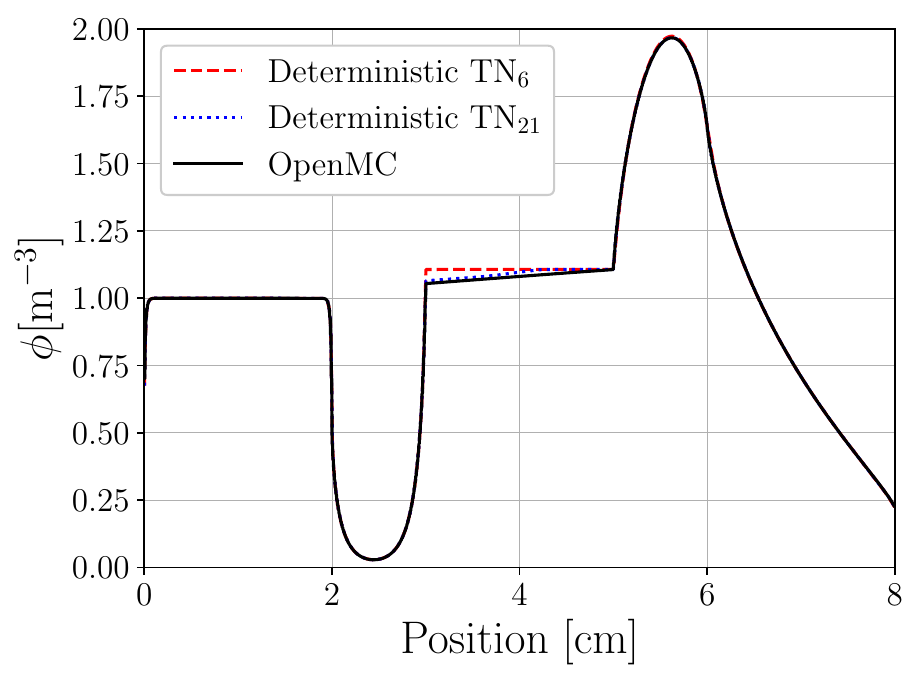}
    \caption{Numerical solutions of the domain setup in figure \ref{fig:reed_setup} for the deterministic code with TN$_6$ quadrature set (dashed red), TN$_{21}$ quadrature set (blue dotted), and OpenMC (black solid) code. The deterministic code uses 20 tetrahedral blocks per cm, and third order elements, while the OpenMC code uses the multigroup mode with 60 billion samples. Agreement is very close (within 2\% in fact), except in region (III) for the TN$_6$ quadrature set, where it is around 5\%. }\label{fig:7_reed_solution}
\end{figure}

The solutions of both the deterministic code (20 element blocks per cm, third order elements, 288 and 3528 angles with the TN$_6$ and TN$_{21}$ quadrature sets) and the OpenMC code (60 billion samples) are shown in figure \ref{fig:7_reed_solution}. The number of elements and order was chosen such that the quadarature set limits the accuracy here. Agreement is within 2\% for both codes, except in region (III) for the TN$_6$ quadrature set, where it is within 5\%. Due to the 3-dimensional geometry, flux can escape on the top and bottom through this region with no cross section. However, the discrete angles that would resolve this effect only enter in the TN$_9$ quadrature set, explaining the flat result in the deterministic code and the observed difference. Furthermore, statistical noise in the OpenMC solution limits the accuracy in some regions, especially region (II).

\begin{figure*}
    \centering
    \includegraphics[width=\linewidth]{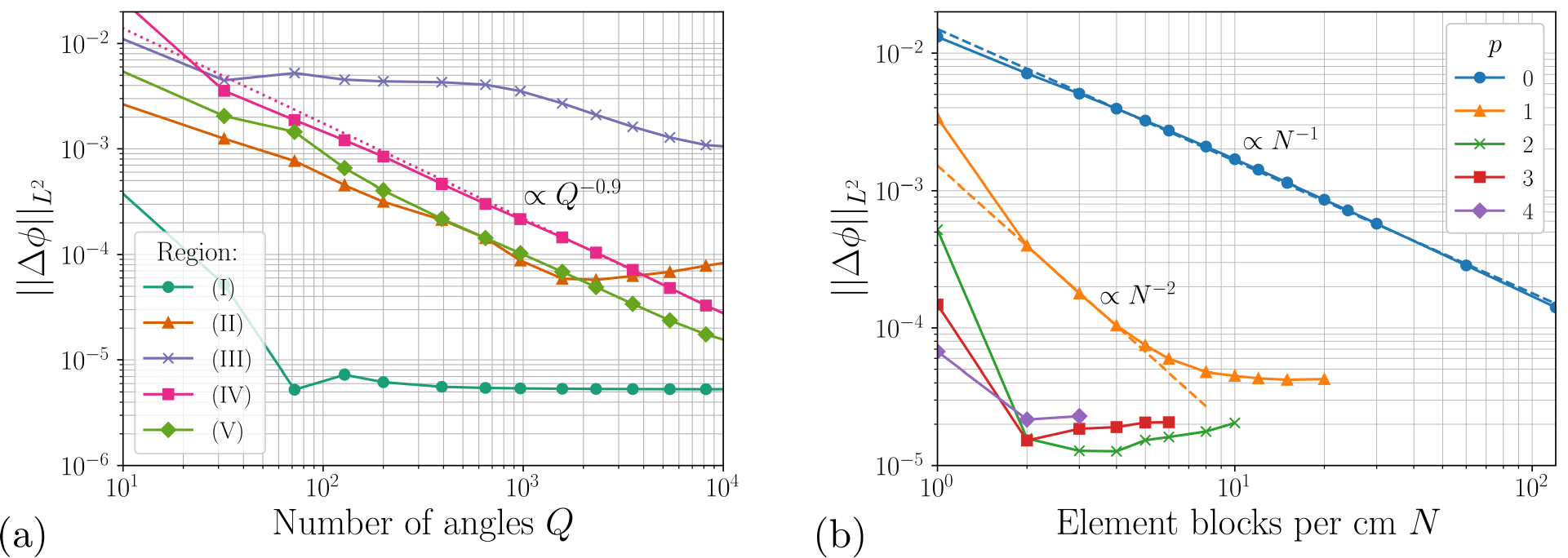}
    \caption{$L^2$ difference of the flux $\phi$ between OpenMC and the deterministic code for the isotropic scattering test as a function of the number of angles for the different regions (a) and as a function of the number of elements per cm (b). For the quadrature convergence, 20 third-order element blocks per cm are used for the spatial discretization. For the spatial convergence, 10952 angles are used for the angular discretization and only regions (IV) and (V) are considered as the other regions are not limited by the spatial discretization but instead by OpenMC or quadrature (region (III)).  Fits are indicated by dashed lines. The quadrature convergence is proportional to $Q^{-0.9}$, slightly less than found in the analytical solution of section \ref{sec:analytical}. The spatial convergence is proportional to $N^{-p+1}$ for orders $p=0,1$, while higher orders do not show a convergence behaviour, as they are limited by OpenMC noise at only 2 element blocks per cm. Note that one element block per cm is the minimum spatial resolution compatible with the mimum feature size of the domain. Also, statistical noise is in fact less for less element blocks per cm, which is why e.g. the $L^2$ difference of second order elements increases with increasing element blocks after 3 element blocks per cm.}\label{fig:7_reed_total_conv}
\end{figure*}

In figure \ref{fig:7_reed_total_conv} the $L^2$ error in the flux is shown as a function of the number of angles $Q$ and number of element blocks per cm $N$. Quadrature convergence rate is here region dependent: in some regions convergence rates close to the analytical problem of section \ref{sec:analytical} are obtained, while other regions such as (I), and with higher number of angles also (II), are not limited by the quadrature set but instead by OpenMC noise and thus show little decreased error with increasing $Q$. Spatial convergence rate is the same as in section \ref{sec:analytical} for lower orders, while higher order elements are limited by OpenMC noise. Note that reducing this statistical noise by an order of magnitude to further investigate these convergence rates would correspond to a hundred times more samples, which proved unfeasible. Here, higher orders are of limited use because accuracy is not severly limited by the spatial discretization due to rather small gradients.

\subsection{Anisotropic multigroup scattering test}\label{sec:anisotropic}

\begin{figure}
    \centering
    \includegraphics[width=\linewidth]{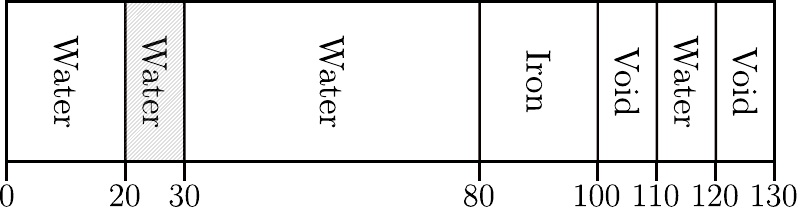}
    \caption{Geometry for the anisotropic multigroup scattering test of references \cite{goldberg2022validating, asaoka1978benchmark, quah1981comparison}. The precise scattering data and matrices are given in \ref{app:1_scatt_data}. The shaded area denotes an isotropic source with strength $[9.7702 \cdot 10^{4}, 4.5451 \cdot 10^5]$ m$^{-3}$s$^{-1}$ respectively for the two energy groups. The cumulative size of each region is shown in cm.}
    \label{fig:8_anisotropic_geometry}
\end{figure}

The anisotropic scattering test is based on \cite{goldberg2022validating, asaoka1978benchmark, quah1981comparison}, with new reference data generated using the multigroup mode of OpenMC. It is a 2-group, one-dimensional test, with the domain is shown in figure \ref{fig:8_anisotropic_geometry}. The specific material scattering data is given in \ref{app:1_scatt_data}. 

\begin{figure*}
    \centering
    \includegraphics[width=\linewidth]{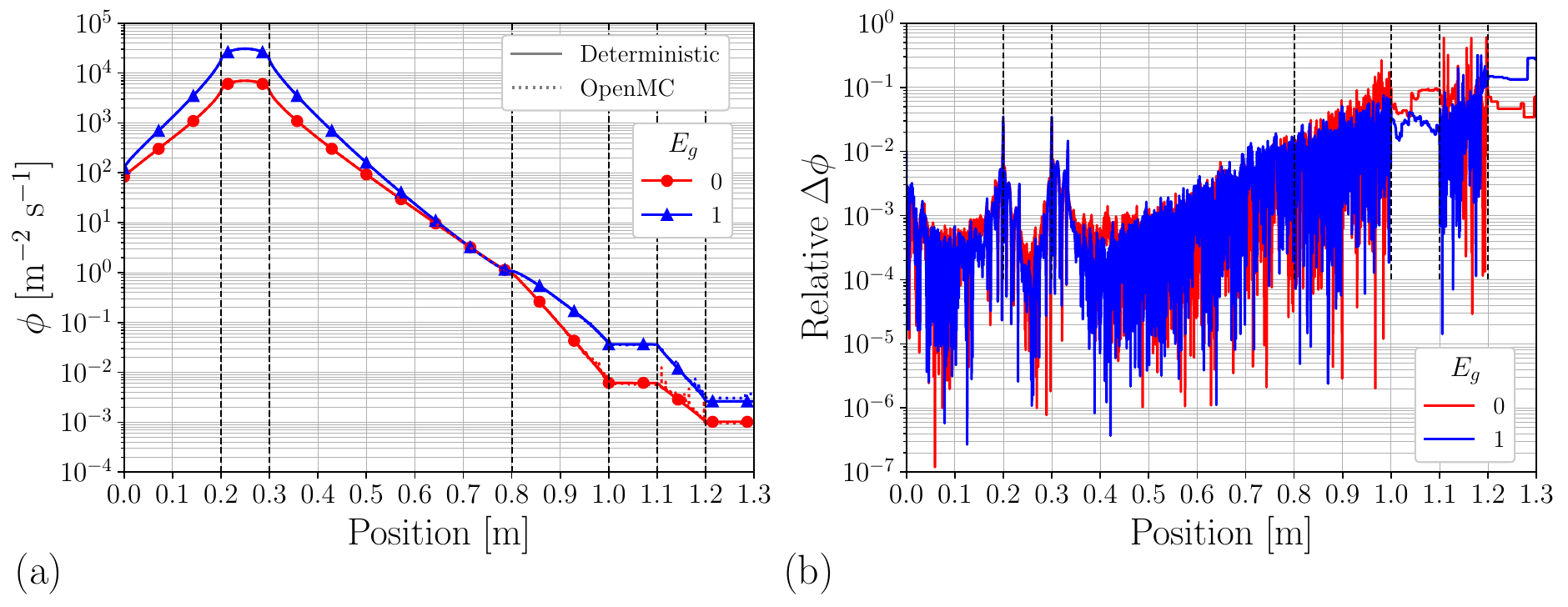}
    \caption{(a) Results for deterministic simulation (solid lines) and an OpenMC simulation (dashed lines) for both energy groups ($E_0$ red dots, $E_1$ blue triangles) for the anisotropic scattering test on the domain of figure \ref{fig:8_anisotropic_geometry}. Black dashed lines indicate the region boundaries of figure \ref{fig:8_anisotropic_geometry}. The deterministic code uses 36 third order elements in the inhomogeneity direction and the TN$_{13}$ quadrature set. The OpenMC simulation uses 10 billion samples. Note the close agreement over seven orders of magnitude. (b) Relative difference between OpenMC and the deterministic simulation. From $x>0.7$, this difference starts to become dominated by OpenMC noise.}
    \label{fig:9_anisotropic_example}
\end{figure*}

The result from a simulation with 36 third order elements (3 per 10 cm) and the TN$_{13}$ quadrature set is shown in figure \ref{fig:9_anisotropic_example} together with an OpenMC simulation (10 billion samples). Close agreement (within 2\% until $x > 0.8$ where OpenMC noise is dominant) is obtained over 7 orders of magnitude.

\begin{figure*}
    \centering
    \includegraphics[width=\linewidth]{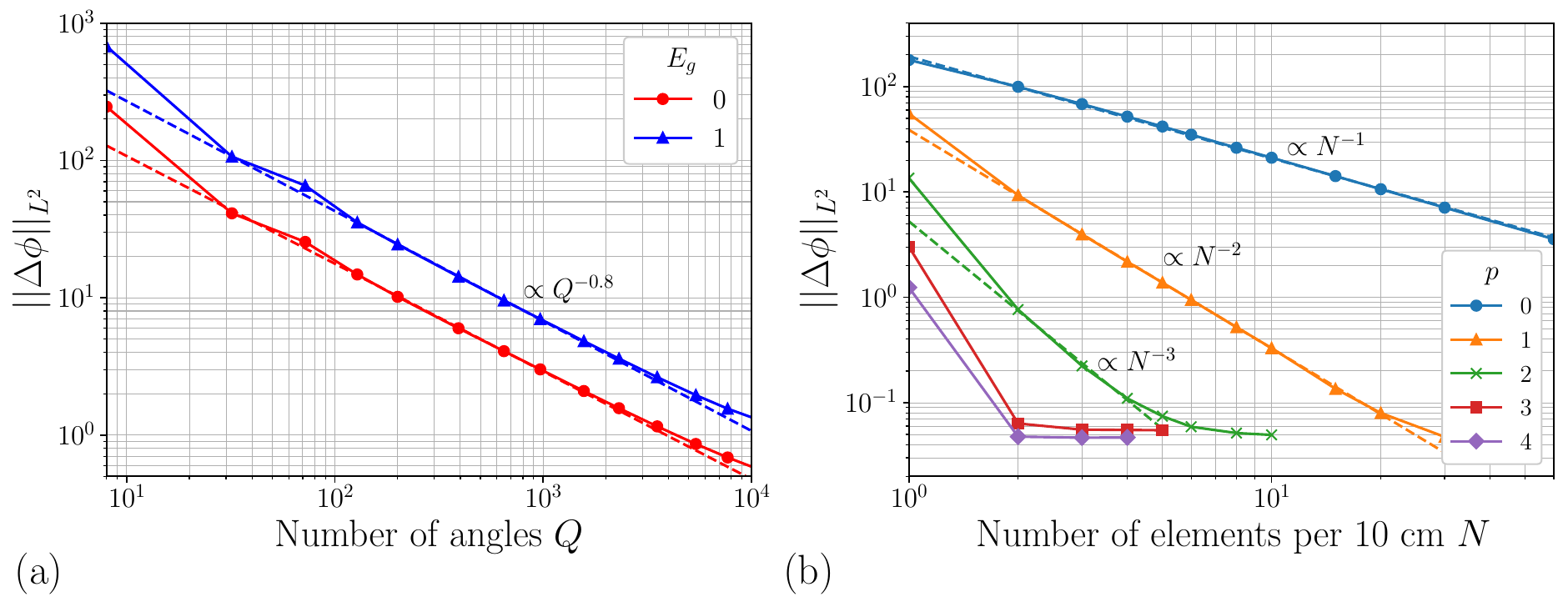}
    \caption{$L^2$ difference of the flux $\phi$ between OpenMC and the deterministic code as a function of the number of angles $Q$ (a) and as a function of the number of elements per 10 cm $N$. Fits are shown with dashed lines. For the quadrature convergence, 52 second order elements per 10 cm are used. For the spatial convergence, the TN13 quadrature set is used and the domain is truncated to $x > 0.4$ because the spatial discretization is the limiting factor there. Quadrature convergence is $Q^{-0.8}$, slightly less than found in the analytical problem of section \ref{sec:analytical}. Spatial convergence is $N^{-p+1}$ as expected for $p=0,1,2$. Higher order elements immediately reach the noise floor from OpenMC, such that no convergence rate can be seen.}\label{fig:10_anisotropic_convergence.pdf}
\end{figure*}

Convergence in both number of quadrature angles and number of spatial elements is shown in figure \ref{fig:10_anisotropic_convergence.pdf}. 52 second order elements are used for quadrature convergence while the TN$_{13}$ quadrature set is used for the spatial convergence. The domain is truncated to $x > 0.4$ because the spatial discretization is the limiting factor there (note the rapid decrease in flux in this region). In this problem, third order elements can reduce the number of elements by a factor of ten for e.g. an $L^2$ error of $10^{-1}$.
Quadrature convergence is $Q^{-0.8}$, again slightly less than the analytical problem in section \ref{sec:analytical}. Spatial convergence rates are equal to the $N^{-p+1}$ of section \ref{sec:analytical}, but can only be seen for $p=0,1,2$ due to the statistical noise of the OpenMC simulation. 

The results in these sections show that the angular and spatial discretization methods are implemented correctly, with the optimal spatial convergence rates achieved in all problems, even in the anisotropic scattering multigroup problem. Furthermore, higher order methods can in some cases significantly reduce the number of elements required in each dimension, decreasing computational requirements, however, in other cases, spatial discretization does not limit the accuracy and thus they are not as useful.


\section{Application to breeding blanket}\label{sec:breedingblanket}
In this section, the full, anisotropic, multigroup deterministic method is tested against the continuous energy OpenMC code with realistic materials relevant for fusion research. In this case, the codes do not solve the same equations, as upscatter is included and the multigroup approximation not made in the OpenMC results. Therefore, this is a relevant test case for the deterministic method. Again, the domain is a three-dimensional slab with one inhomogeneity direction. 

In this inhomogeneity direction, the first meter is a source region with a constant cross section of $\sigma_t = 10$ m$^{-1}$ (needed to make the solution homogeneous in the other two directions). It isotropically emits $14.1$ MeV neutrons with a normalized rate of 1 m$^{-3}$ s$^{-1}$. This is attached to a layered breeding blanket based on \cite{warmer2017w7}. The thickness of each region is shown in table \ref{tab:breedingblanket_thickness}. Material compositions of each named layer are shown in table \ref{tab:breedingblanket_materials}. The homogeneous dimensions are 10m long.

\begin{table}[ht]
\caption{\label{tab:breedingblanket_thickness} Thickness of each region in the one-dimensional breeding blanket model. Thicknesses based on \cite{warmer2017w7}.}
\lineup
\begin{indented}
\item[]\begin{tabular}{@{}l|l|l}
\br
 \textbf{Region} & \textbf{Thickness} [m]&\textbf{Cumulative} [m]\\
\hline
    Tungsten Armor & 0.002 & 0.002\\ 
    First Wall & 0.025 & 0.027 \\ 
    Breeding Zone & 0.5 & 0.527 \\ 
    Back Support Structure & 0.385 & 0.912 \\ 
    Vacuum Vessel Wall & 0.06 & 0.972 \\ 
    Vacuum Vessel Shield& 0.20 & 1.172 \\ 
    Vacuum Vessel Wall & 0.06 & 1.232 \\ 

\end{tabular}
\end{indented}
\end{table}

\begin{table}[ht]
\caption{\label{tab:breedingblanket_materials} Material composition of each named region in table \ref{tab:breedingblanket_thickness}. Based on \cite{lyytinen2023parametric}.}
\lineup
\begin{indented}
\item[]\begin{tabular}{@{}l|l|l}
\br
 \textbf{Region} & \textbf{Material} &\textbf{Volume \%}\\
\hline
    Tungsten Armor& W$^{nat}$ & 100\%\\
    \hline
    First Wall & He$^{nat}$ & 35.09\%\\ 
               & Eurofer & 64.91\%\\ \hline
    Breeding Zone &      Eurofer & 9.85\%\\ 
                  &     He$^{nat}$ & 38.61\%\\ 
                  & Li$_4$SiO$_4$ (60\% Li$_6$) & 14.91\%\\ 
                  & Be$^{nat}$ & 36.63\% \\ \hline 
    
    Back Support Structure   &      Eurofer & 60.61\%\\ 
                                                & He$^{nat}$ & 39.39\%\\ \hline
    Vacuum Vessel Wall  & SS316LN & 100\% \\ \hline
    Vacuum Vessel Shield & SS316LN & 60.0\% \\
                                     & Water & 40.0 \% \\ \hline

\end{tabular}
\end{indented}
\end{table}
\subsection{OpenMC setup}
The OpenMC code will only be run in continuous energy mode, as the multigroup mode does not work with the highly anisotropic scattering used in these problems due to negative scattering probabilities \cite{Martin2011}. The constant cross section of the source region is obtained by modifying the source code to include a dummy nuclide that has a total and absorption cross section of 10 irrespective of the neutron energy. 

The mulitgroup cross section generation capability of OpenMC \cite{boyd2019multigroup} is used to generate multigroup cross sections on the Vitamin-J-175 group structure \cite{sartori1985vitamin}. These can optionally be used by the deterministic code. 

Results are tallied in the inhomogeneity direction using a 1mm fine mesh to compare with the results from the deterministic code. Three billion samples are used.
\subsection{Deterministic code setup}
Third order elements are used with a resolution of (4,1,1,6,4,2,2,2) elements in the respective region. The TN$_5$ quadrature set is used. The other dimensions use 3 elements. 

Cross sections are obtained in two ways. First, using TRANSX \cite{macfarlane1992transx} on the FENDL3.2b \cite{schnabel2024fendl} dataset. This dataset uses the aforementioned Vitamin-J 175 group structure and $P_5$ ($L=5$) anisotropy. Only the 168 energy groups below 14.1 MeV are calculated, as there is no upscatter above 14.1 MeV. Upscatter cross sections are generated but are identically zero for $E > 10$ eV, while being below $5\%$ of the total cross section for $E < 10$ eV. As mentioned in section \ref{sec:assumptions}, they are however not used in the deterministic code (as upscatter is neglected). This will only impact the energy range $E<10$ eV, as there is thus no upscatter for $E > 10 $ eV.

The second method uses the generated OpenMC multigroup cross sections. These are generated for the specific problem and are thus expected to perform significantly better.

\subsection{Comparison}
\begin{figure*}
    \centering
    \includegraphics[width=\linewidth]{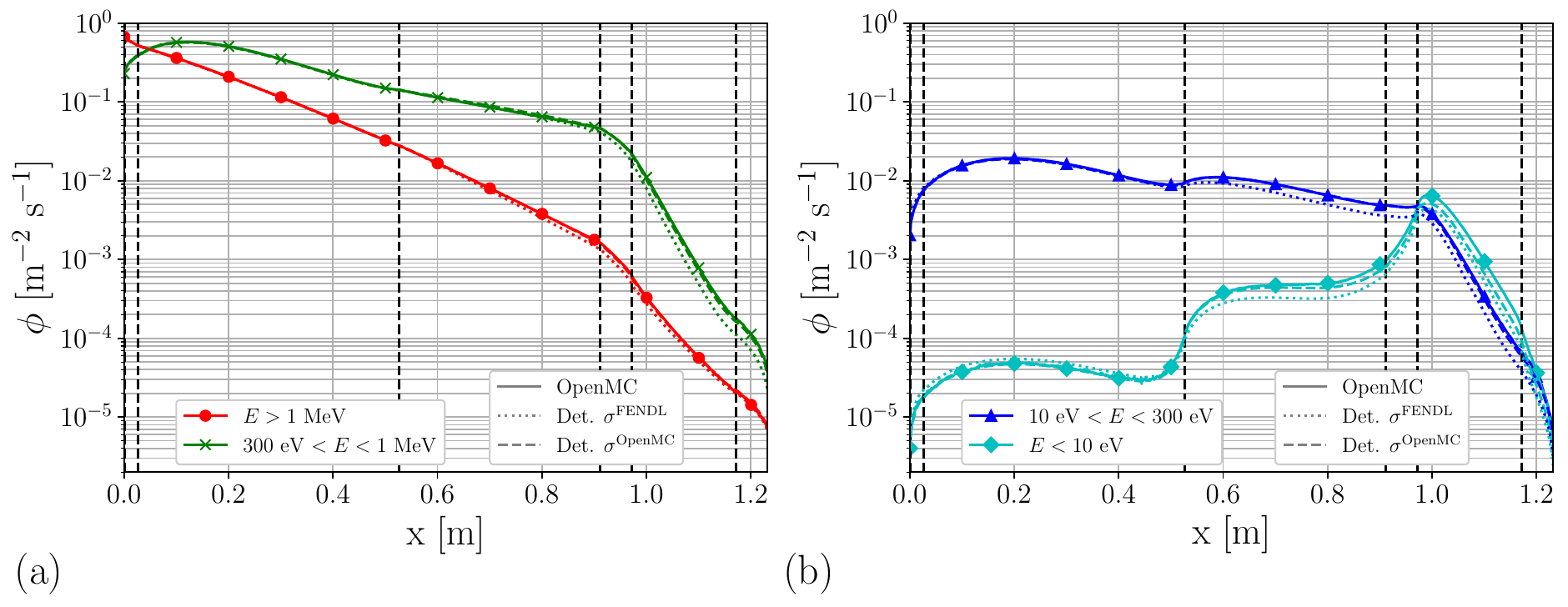}
    \caption{Scalar flux as a function of position in the breeding blanket as defined in tables \ref{tab:breedingblanket_materials} and \ref{tab:breedingblanket_thickness} for the OpenMC simulation (solid lines), and the deterministic code with FENDL cross sections (dotted lines) and the OpenMC generated cross sections (dashed lines). The scalar flux is shown for four different energy ranges: $E>1$ MeV (red dots), $300$ eV $< E <$ 1 MeV (green crosses) in figure (a), and $10$ eV $ < E < 300$ eV (blue triangles) and $E<10$ eV (cyan diamonds) in figure (b).  Simulations are in good agreement across all energy ranges and positions over five orders of magnitude, although for the lower energy ranges the simulations differ slightly, especially for the FENDL cross sections.}\label{fig:11_scalarbb_comparison}
\end{figure*}

The results of the OpenMC simulation and the deterministic simulations with the FENDL3.2b and OpenMC-generated cross sections are shown in figure 
\ref{fig:11_scalarbb_comparison}. Agreement is qualitatively good across the whole blanket and all energy ranges, although for lower energies at the edge of the blanket, results differ, especially for the FENDL cross sections. For the FENDL cross sections, agreement is within 10\% in the breeding zone for $E>10$ eV but agreement reduces towards the edge of the blanket. For the OpenMC generated cross sections, agreement is within 10\% for $E>10$ eV for the entire blanket.

\begin{figure*}
    \centering
    \includegraphics[width=\linewidth]{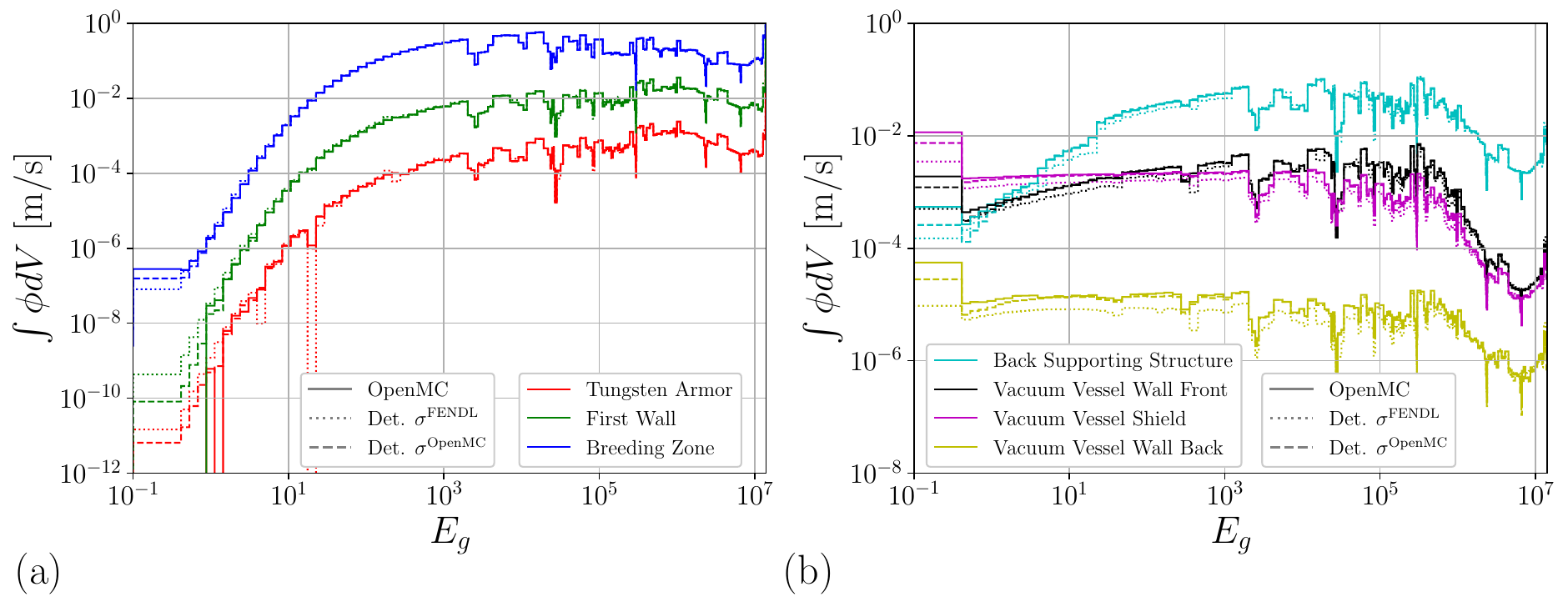}
    \caption{Scalar flux spectrum integrated over the regions in the breeding blanket as defined in tables \ref{tab:breedingblanket_thickness}, \ref{tab:breedingblanket_materials} and for the OpenMC simulation (solid lines), and the deterministic code with FENDL cross sections (dotted lines) and the OpenMC generated cross sections (dashed lines). Simulations are in qualitative agreement across the entire energy spectrum over 12 orders of magnitude. Especially in the lower energy range, OpenMC suffers from statistical noise (which also impacts the cross sections generated from OpenMC).}\label{fig:12_scalarspectrum_comparison}
\end{figure*}

The energy spectra volume integrated over each region are shown in figure \ref{fig:12_scalarspectrum_comparison}. Agreement is good across the entire energy range and for each region, although for lower energy ranges in the edge of the blanket differences are observed, especially for the FENDL cross sections.

Another relevant quantity for fusion research is the total tritium breeding. Since the tritium breeding rate is simply the scalar flux times the tritium breeding cross section, a close agreement of the scalar flux implies a close agreement of the tritium breeding rate as long as the multigroup cross section is accurate. In these cases, tritium breeding rate is only 1.5\% lower for the FENDL cross sections and 1\% lower for the OpenMC generated cross sections (note that differences in the edge of the blanket do not impact the tritium breeding rate due to a lack of breeding material there).

The observed differences are not due to a lack of sufficient spatial or angular resolution; they arise due to the limitations of the multigroup and Legendre polynomial based scattering source approximations. Increasing spatial and angular resolution significantly beyond the results shown here (up to TN30 quadrature set with 640 third order elements in the $x$-direction was tested in 1D, thus also removing edge effects completely) does not result in better agreement: all results in the edge of the blanket still only agree within 10\%. Furthermore, the cross sections specifically generated for this problem by OpenMC significantly improve the results without changing the spatial or angular resolution. In any case, the FENDL cross sections produce qualitatively very similar results and important results such as tritium breeding rate and fast neutron flux at the blanket edge agree within  10\%. 

\begin{figure}
    \centering
    \includegraphics[width=\linewidth]{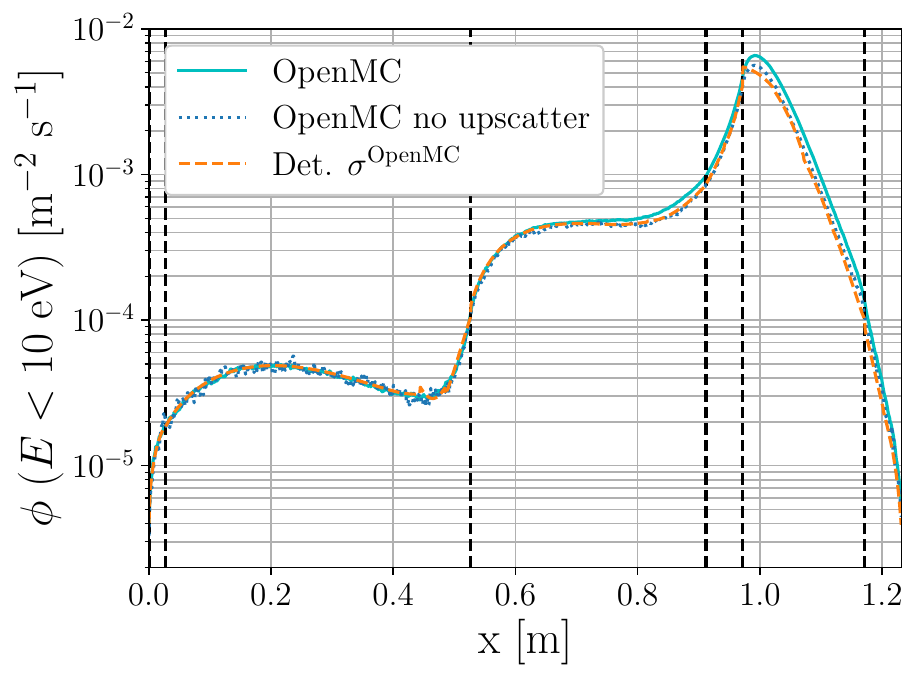}
    \caption{Scalar flux in the energy range $E < 10$ eV where upscatter is significant for the OpenMC simulation (cyan solid line, shown also in figure \ref{fig:11_scalarbb_comparison}), an OpenMC simulation where group-to-group upscatter was manually removed (blue dotted line) and the deterministic simulation with OpenMC-generated cross sections (orange dashed line, also shown in figure \ref{fig:11_scalarbb_comparison}). The upscatter assumption partly explains the observed differences in figure \ref{fig:11_scalarbb_comparison}.}
    \label{fig:13_upscatter}
\end{figure}

Finally, although the upscatter assumption affects the energy range $E < 10 $ eV, results still agree qualitatively quite well. As can be seen in the spectra of figure \ref{fig:12_scalarspectrum_comparison}, the last few energy groups ($E < 1$ eV) are the main source of error. This error results from not only the upscatter assumption, but also from the multigroup approximation itself (the group structure considered is not optimized particularly for low-energy or thermal computations \cite{fleming2016optimization}). In figure \ref{fig:13_upscatter}, an OpenMC simulation with the group-to-group upscatter effect removed is compared to the simulations in figure \ref{fig:11_scalarbb_comparison}, showing that this assumption partly explains the observed differences for $E<10$ eV (for higher energy groups, there is no difference between the OpenMC simulations with or without upscatter).

\subsection{Breeding Blanket with 3D inhomogeneity}\label{sec:inhom_blanket}
\begin{figure}
    \centering
    \includegraphics[width=\linewidth]{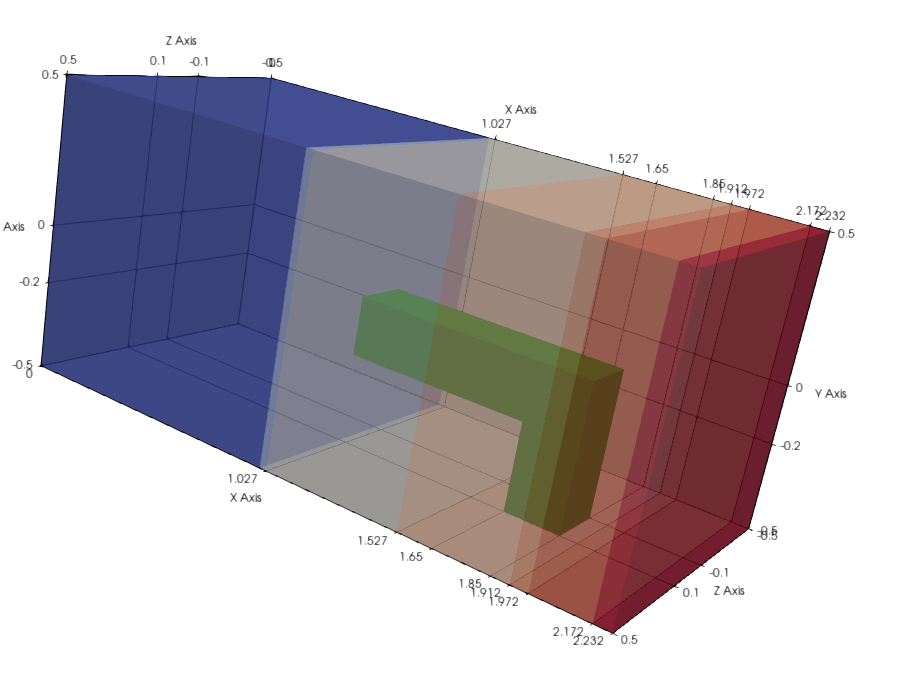}
    \caption{The three-dimensional blanket geometry of section \ref{sec:inhom_blanket}. The color indicates the material and the material boundaries are indicated on the $x$-axis (see table \ref{tab:breedingblanket_thickness}). The cutout containing the void material is shown in green, with its boundaries also shown on the $x,y,z$ axes.}
    \label{fig:14_dogleg_figure}
\end{figure}

In this subsection, a three-dimensional inhomogeneity is introduced. This thus becomes a fully 3D benchmark, albeit in relatively simple geometry. First, the size in the $y,z$ dimensions is reduced to 1.0 meter to re-introduce edge effects, with the center of the $yz$-cross section being set at $(0,0)$. Furthermore, in the blanket itself, a region is cut out and filled with void material. This region is comprised of two rectangular beams: 
\begin{eqnarray*}
    x_1 &= \{1.00, 1.85\},\     y_1 = \{-0.2, 0.0\},\      z_1 = \{-0.1, 0.1\}\\
    x_2 &= \{1.65, 1.85\},\     y_2 = \{-0.5, 0.0\},\      z_2 = \{-0.1, 0.1\}
\end{eqnarray*}
The resulting geometry is shown in figure \ref{fig:14_dogleg_figure}. Only OpenMC-generated cross sections (from the 1D problem) are used. The $TN_9$ quadrature set was used to limit ray-effects in the cutout. The resolution was slightly increased in the $x$-direction (the back-supporting structure now has 10 blocks instead of 4 to accommodate the cutout), and in the $y,z$ directions the number of blocks was increased to 19 and 17 respectively. The OpenMC simulation uses 6 billion samples.

\begin{figure*}
    \centering
    \includegraphics[width=\linewidth]{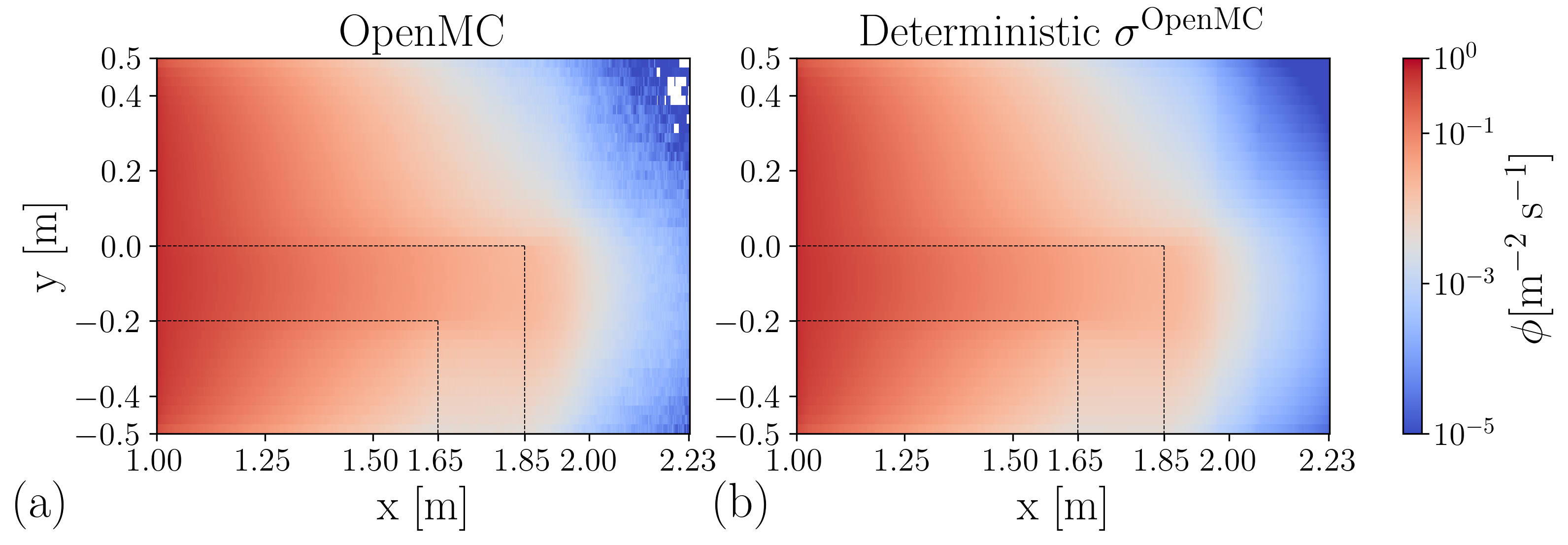}
    \caption{Scalar flux for $E>1 \ MeV$ averaged over the cutout in the $z$-dimension for the geometry of figure \ref{fig:14_dogleg_figure}. Figure (a) shows the OpenMC results, and figure (b) the deterministic results with the cross sections from the 1D OpenMC simulation. The cutout in the $x,y$ dimensions is shown with black dashed lines. Qualitatively, the simulations are very similar. Statistical noise is present in the edge of the blanket (compared to the 1D simulations, this simulation requires a full 3D resolution of the distribution function and thus increased sampling). }
    \label{fig:15_2D_scalar_flux}
\end{figure*}

\begin{figure*}
    \centering
    \includegraphics[width=\linewidth]{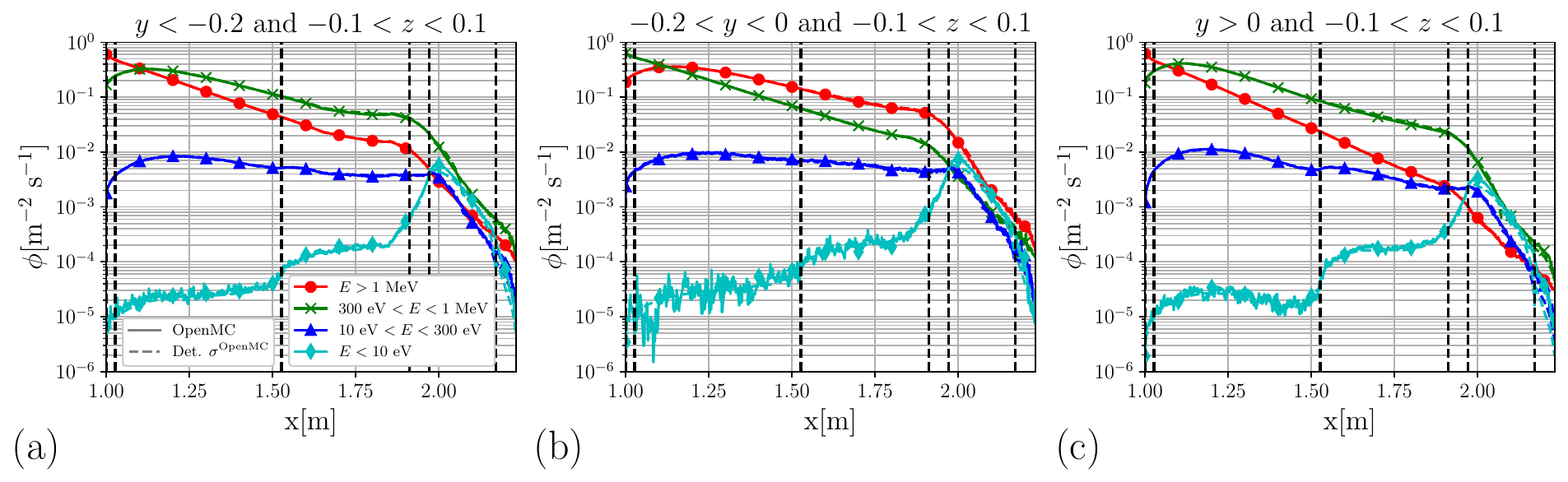}
    \caption{Scalar flux in the geometry of figure \ref{fig:14_dogleg_figure} as a function of $x$, averaged over the cutout in the $z$-dimension ($-0.1 < z<0.10$) and over the region below the cutout ((a), $y < -0.2$), over the cutout itself ((b), $-0.2 < y <0$) and over the region above the cutout ((c), $y>0$). Solid lines denote the OpenMC solution, dashed lines the deterministic solution using cross sections from the 1D OpenMC simulation. Red circles denote $E > 1 \ MeV$, green crosses $300 \ eV < E < 1 \ MeV$, blue triangles $10 \ eV < E < 300 \ eV$, and cyan diamonds $E< 10 \ eV$. Quantitative agreement is good across all energy ranges and all regions of the blanket, except at the edge for the low energy ranges.}
    \label{fig:16_1D_scalar_cutout}
\end{figure*}

A two-dimensional scalar flux comparison for $E>1 \ MeV$ averaged over the cutout in the $z$-dimension is shown in figure  \ref{fig:15_2D_scalar_flux}, showing qualitative agreement. 
To compare quantitatively, the flux is, after being averaged over the cutout, also averaged over three regions: $y < -0.2$, $-0.2  < y < 0$, and $ y > 0$, corresponding to the regions in figure \ref{fig:15_2D_scalar_flux} under the cutout, in the cutout, and above the cutout. The results for these three regions can be seen in figure \ref{fig:16_1D_scalar_cutout}, showing quantitative agreement in all regions and across all energy ranges. Agreement is within 10
\%, except in the edge of the blanket for the lower energy ranges. Statistical noise is also present in the OpenMC solution, however, especially in the edge and for the lower energy range.
The results in this section show that the deterministic code can produce accurate results in a fusion-relevant context, even in three-dimensional geometry. Furthermore, although a detailed and fair comparison of computational efficiency is out of the scope of this work, the results for the deterministic code use significantly less time CPU-hours in all considered cases, even though this simple planar geometry is advantageous for the Monte-Carlo method (and in the three-dimensional case, the solution is not fully converged either). Finally, generating problem-specific cross sections can improve the deterministic solution significantly over the general FENDL cross sections, showing good results even in a modified geometry. In a design-optimization context, where the design is updated only slightly (e.g. a slightly increased breeding zone thickness), it can therefore be useful to generate these cross section once and use the deterministic code to investigate the result of such a slight change.  


\section{Conclusion}\label{sec:conclusion}
Neutronics are a key design driver for fusion reactors, such that a fast and accurate evaluation of neutronic parameters is of paramount importance. Conventionally, either reduced models or statistical methods are used for calculating them. However, in a design context, and particularly for early-stage design and optimization, neither is ideal: reduced models provide too little information or accuracy to properly inform the three-dimensional design, while statistical methods take too long for a fast design cycle. 

In this work, a method was presented that could provide a middle-ground between these extremes. It uses a novel combination of higher order Discontinuous-Galerkin on unstructured meshes, matrix-free iterative solvers, and the multigroup, discrete ordinates phase space discretization to solve the neutron transport equation in three-dimensional geometry. A detailed derivation of the scheme was given, including a significant simplification for simplexes and the implementation of the matrix-free iterative solvers. A novel approach to parallelization for higher order elements was also explained. 

The resulting model was then verified in slab geometry using an analytical solution, showing the expected higher-order spatial convergence rates. Then, two literature tests with isotropic and anisotropic multigroup scattering were performed, using a multigroup Monte-Carlo code to generate reference solutions. These tests have shown that in the presence of scattering the model still exhibits the higher order convergence rates. 

Finally, a fusion relevant helium-cooled-pebble bed breeding blanket was simulated using both the deterministic model and a conventional Monte-Carlo code. This demonstrated that the deterministic method can produce results very close to the detailed Monte-Carlo analysis for a fusion relevant problem, both in one and three dimensions.

These results indicate that the deterministic method can in principle be used to asses the required neutronic parameters accurately. However, before using it in reactor design or optimization, it should be benchmarked with established codes in a more complex, three-dimensional reactor geometry, including computational speed comparisons. This is the subject of a future paper. 

Further possible future extensions include photon transport, coupling to thermo-mechanical engineering codes, curvilinear angular discretizations for e.g. 2D tokamak analyses, GPU acceleration, differentiable solutions, more physical thermal neutron treatment, or coupling to Monte-Carlo codes for variance reduction.
\section{Acknowledgements}
This work has been carried out within the framework of the EUROfusion Consortium, funded by the European Union via the Euratom Research and Training Programme (Grant Agreement No 101052200 — EUROfusion). Views and opinions expressed are, however, those of the author(s) only and do not necessarily reflect those of the European Union or the European Commission. Neither the European Union nor the European Commission can be held responsible for them.
\section*{References}
\bibliography{iopart-num}

\appendix
\section{Derivation of anisotropic scattering source}\label{app:deriv_scat_source}
Although this is a standard practice, the derivation using the same notation and normalization is given here for completeness. See \cite{Lewis1984} for a more complete derivation. The total scattering source can be written as 
\begin{equation}
    q^s(\mathbf{r}, E, \hat{\bOmega}) = \int\int \sigma^s(\mathbf{r}, \tilde{\bOmega}\cdot \bOmega, \tilde{E}\to E) \psi d\tilde{\bOmega}d\tilde{E} 
\end{equation}
where $\sigma^s$ is the differential scattering cross section and $\psi = \psi(\mathbf{r}, \tilde{E}, \tilde{\bOmega})$. Using the multigroup approximation, the energy integral reduces to a sum as 
\begin{equation}\label{eq:app_eg_scat}
    q^s_g(\mathbf{r}, E ,\hat{\bOmega}) = \sum_{\tilde{g}=1}^G \int\sigma^s(\mathbf{r},  \tilde{\bOmega}\cdot \homega, E_{\tilde{g}}\to E_g) \psi_{\tilde{g}} d\tilde{\bOmega}
\end{equation}
The scattering cross section expansion in Legendre polynomials $P_l$ with maximum order truncated to $L$ is\footnote{The $(2l+1)$ factor is not present in all codes and in some cases is absorbed into $\sigma^l_{g\tilde{g}}$}
\begin{equation}
    \sigma^s(\mathbf{r},  \tilde{\bOmega}\cdot \hat{\bOmega}, E_{\tilde{g}}\to E_g) \approx \sum_{l=0}^L (2l + 1) P_l(\tilde{\bOmega}\cdot \hat{\bOmega} ) \sigma^l_{g\tilde{g}}(\mathbf{r}),
\end{equation}
where $\sigma^l_{g\tilde{g}}$ is the expansion coefficient. As the Legendre polynomials are complete, this converges to the actual $\sigma^s$ as $L\to\infty$.
Filling this into equation \eref{eq:app_eg_scat} yields
\begin{equation}\label{eq:app_scat_source_ex}
    q^s_g = \sum_{\tilde{g}=1}^G \sum_{l=0}^L \int\sigma^l_{g\tilde{g}}  (2l +1)P_l( \tilde{\bOmega}\cdot \homega)\psi_g d\tilde{\bOmega}.
\end{equation}
Now, the spherical harmonics used in this work are defined as (note the normalization is different than in some other definitions by a factor $\sqrt{4\pi}$)
\begin{equation}
    Y_{lm}(\hat{\bOmega}(\theta, \phi)) = \sqrt{(2l+1) \frac{(l-m)!}{(l+m)!}}P_l(cos(\theta)) e^{im\phi}
\end{equation}
Furthermore, define
\begin{equation}
    \int d\bOmega = 1,
\end{equation}
then,
\begin{equation}\label{app:ortho}
    \int Y^*_{lm} Y_{\lambda \mu}d\bOmega = \delta_{l\lambda}\delta_{m\mu}
\end{equation}
where $\delta$ is the Kronecker delta. As the spherical harmonics are complete on the unit sphere, $\psi$ can be expanded as 
\begin{eqnarray}
    \psi(\mathbf{r}, E_{\tilde{g}}, \tilde{\bOmega}) &= \sum_{\lambda=0}^\infty\sum_{\mu=-\lambda}^{\lambda}  \phi_{\lambda \mu}(\mathbf{r}, E_{\tilde{g}})Y^*_{\lambda \mu}(\tilde{\bOmega}) \\
    :&= \sum_{\lambda,\mu}  \phi_{\lambda \mu}(\mathbf{r}, E_{\tilde{g}})Y^*_{\lambda \mu}(\tilde{\bOmega})
\end{eqnarray}

with 
\begin{equation}
    \phi_{\lambda \mu}(\mathbf{r}, E_{\tilde{g}}) :=\phi^{\tilde{g}}_{\lambda \mu}  = \int\psi_{\tilde{g}} Y_{\lambda \mu}(\hat{\bOmega})d \hat{\bOmega}   .
\end{equation}
Equation \eref{eq:app_scat_source_ex} is then 
\begin{equation}\label{eq:app_both_ex}
    \sum_{\tilde{g}=1}^G \sum_{l=0}^L \int\sigma^l_{g\tilde{g}}  (2l+1) P_l( \tilde{\bOmega}\cdot \homega)   \sum_{\lambda,\mu}  \phi^{\tilde{g}}_{\lambda \mu}Y^*_{\lambda \mu}(\tilde{\bOmega})  d\tilde{\bOmega}
\end{equation}
Taking all sums and other coefficients out of the integral yields 
\begin{equation}\label{eq:app_both_ex_sum}
    \sum_{\tilde{g}=1}^G \sum_{l=0}^L  \sum_{\lambda,\mu}  \sigma^l_{g\tilde{g}}  \int (2l+1) P_l( \tilde{\bOmega}\cdot \homega)    \phi^{\tilde{g}}_{\lambda \mu}Y^*_{\lambda \mu}(\tilde{\bOmega})  d\tilde{\bOmega}
\end{equation}
Then, the Legendre addition theorem (with the normalization used here for the spherical harmonics):
\begin{equation}
    (2l+1)P_l(\tilde{\bOmega}\cdot \hat{\bOmega}) = \sum_{m=-l}^l Y_{lm}(\tilde{\bOmega})Y^*_{lm}(\hat{\bOmega})
\end{equation}
can be filled into equation \eref{eq:app_both_ex_sum} as 
\begin{equation}\label{eq:app_thrice_ex_sum}
    \sum_{\tilde{g}=1}^G \sum_{l,m} \sum_{\lambda,\mu}  \sigma^l_{g\tilde{g}} \phi^{\tilde{g}}_{\lambda \mu} Y^*_{lm}(\hat{\bOmega}) \int  Y_{lm}(\tilde{\bOmega}) Y^*_{\lambda \mu}(\tilde{\bOmega})  d\tilde{\bOmega}
\end{equation} 
Using the orgonality of equation \eref{app:ortho} then yields 
\begin{eqnarray}
    &\sum_{\tilde{g}=1}^G \sum_{l=0}^L \sum_{m=-l}^l \sum_{\lambda=0}^\infty\sum_{\mu=-\lambda}^{\lambda}  \sigma^l_{g\tilde{g}} \phi^{\tilde{g}}_{\lambda \mu} Y^*_{lm}(\hat{\bOmega})\delta_{l\lambda}\delta_{m\mu} \nonumber \\ 
    &=\sum_{\tilde{g}=1}^G \sum_{l=0}^L \sum_{m=-l}^l   \sigma^l_{g\tilde{g}} \phi^{\tilde{g}}_{lm}(\mathbf{r}) Y^*_{lm}(\hat{\bOmega})
\end{eqnarray}
Finally, split the energy groups into upscatter ($\tilde{g} > g$), inscatter ($\tilde{g}=g$) and downscatter ($\tilde{g} < g$) and use that upscatter is not present:
\begin{eqnarray*}
    q(\mathbf{r}, E_g, \hat{\bOmega}) &=    \sum_{\tilde{g}=1}^{g-1} \sum_{l=0}^L \sum_{m=-l}^l   \sigma^l_{g\tilde{g}} \phi^{\tilde{g}}_{lm}(\mathbf{r}) Y^*_{lm}(\hat{\bOmega})\\
&+\sigma^l_{gg} \phi^{g}_{lm}(\mathbf{r}) Y^*_{lm}(\hat{\bOmega})\\
&=q^D_{g} + q^I_g
\end{eqnarray*}
with $q^D_g$ being downscatter to group $g$ and $q^I_g$ being inscatter in group $g$.  This is equal to equation \ref{eq:scat_split}.

\section{Operator form derivation}\label{app:matrix_free}
Writing the neutron transport equation for energy group $E_g$ including the indices for both the solution values $j$ and the discrete angles $\hat{\bOmega}_k$ and using the expansion of the scatter source,  yields the set of equations (note that this at the moment still for each element independently)

\begin{eqnarray}
    \fl \sum_j \left[\int_M \sigma H_i H_j dV  - \int_M H_j \hat{\bOmega} \cdot \nabla H_i dV \right.  
\nonumber \\
\left.+ \sum_{e^+} |\hat{\bOmega}\cdot \hat{\mathbf{n}}_e |\int_e H_i H_j dA\right]\psi_{kj} \nonumber \\
    = \sum_{e^-} |\hat{\bOmega}\cdot \hat{\mathbf{n}}_e| \sum_\eta \psi^O_{k\eta}   \int_e H_i \overline{H}_\eta dA \nonumber  \\ + \sum_j \left[q^e_{kj} +\sum_{lm}Y^*_{lmk}\left(q^D_{lmj} + \sigma^l_{gg} \sum_{\kappa} w_\kappa \psi_{\kappa j}Y_{lm\kappa}\right)\right]\nonumber \\ \cdot \int  H_i H_j dV \label{eq:total_equation_indices0} 
\end{eqnarray}
where $Y_{lmk}$ is the spherical harmonic $lm$ evaluated at discrete ordinate $\bOmega_k$.
These equations should hold for all discrete angles $k$ and element basis functions $i$. Then, note that all basis functions $H_i$ not belonging to element $M$ will produce zero overlap integrals in any case, such that the equations should hold for all $i$, i.e. $H_i$ also belonging to other elements. Then, the index $j$ can be taken to be over all elements as well. Finally, define $\overline{H}^O_j$ to be only the basis functions that belong to element $O$ and zero for all others. Then, $\overline{H}_\eta$ can be replaced with $\overline{H}^O_j$ and $\psi_{gk\eta}$ with $\psi_{gkj}$, as only the correct inflow terms will remain. This yields:

\begin{eqnarray}
    \fl \sum_j \left[\int_M \sigma H_i H_j dV  - \int_M H_j \hat{\bOmega} \cdot \nabla H_i dV \right.  
\nonumber \\
\left.+ \sum_{e^+} |\hat{\bOmega}\cdot \hat{\mathbf{n}}_e |\int_e H_i H_j dA - \sum_{e^-} |\hat{\bOmega}\cdot \hat{\mathbf{n}}_e|    \int_e H_i \overline{H}^O_j dA \right]\psi_{kj}\nonumber  \\
-\sum_j\sum_{lm}Y_{lmk} \sigma^l_{gg} \sum_{\kappa} w_\kappa \psi_{\kappa j}Y_{lm\kappa}\int  H_i H_j dV 
 \nonumber \\
    =  \sum_j q^{e,D}_{kj}\int  H_i H_j dV \equiv Q_{ki}\label{eq:total_equation_indices} 
\end{eqnarray}
where $q^{e,D}_{kj}$ is the combined external and downscatter source and $Q_{ki}$ is the resulting total base source term. Now, essentially, two types of values exist: values defined on all element values $i,j$ and all angles $\Omega_k$ (e.g. $\psi_{kj}$), and values defined for all element values $i,j$ and all spherical harmonic moments $lm$ (e.g. $\phi_{lmj}$). These types of values exist within the $\mathcal{D}, \mathcal{M}$ spaces (discrete $\mathcal{D}$, moments $\mathcal{M}$). Then, the following linear operator can be defined that takes values from the $\mathcal{D}$ to the $\mathcal{M}$ space:
\begin{eqnarray}
    \mathbf{M}: \ \ \mathcal{D} \to \mathcal{M};   \ \    \psi_{kj} \to \sum_{k} w_k \psi_{kj} Y_{lmk} := \phi_{lmj}
\end{eqnarray}
Furthermore, the linear (in the sense of mapping, not linear anisotropic scattering) scattering operator can be written as an operator from the angular moments $\phi_{lmj}$  to the inscatter source $q^I_{ki}$ at discrete angle $\bOmega_k$ at element value $i$
\begin{eqnarray}
    \mathbf{S}: \ \ \mathcal{M} \to \mathcal{D}; \ \   \phi_{glmj} \to q^I_{gki} := \nonumber \\
          \ \ \ \ \ \ \ \ \ \ \ \ \sum_j \sum_{lm} Y_{lmk} \sigma^l_{gg} \phi_{glmj} \int H_i H_j dV
\end{eqnarray}
where now 
\begin{eqnarray}
    \mathbf{S} \circ \mathbf{M}: \ \ \mathcal{D} \to \mathcal{D};\ \  \psi_{gkj}\to q^I_{gki} \equiv \nonumber \\ 
    \sum_j\sum_{lm}Y_{lmk} \sigma^l_{gg} \sum_{\kappa} w_\kappa \psi_{g\kappa j}Y_{lm\kappa}\int  H_i H_j dV 
\end{eqnarray}

Finally, the linear transport operator $\mathbf{T}$ takes as argument the solution values $\psi_{gkj}$ and calculates the following integrals and sums with basis function $H_i$:
\begin{eqnarray}
    \mathbf{T}: \ \ \mathcal{D} \to \mathcal{D}; \ \ &\psi_{gkj} \to T_{gki}:= \nonumber \\
\fl \sum_j \left[\int_M \right. \sigma_g H_i H_j &dV - \left.\int_M H_j \hat{\bOmega} \cdot \nabla H_i dV \right.  
\nonumber \\
&+ \sum_{e^+} |\hat{\bOmega}\cdot \hat{\mathbf{n}}_e |\int_e H_i H_j dA \nonumber\\ 
 &\left. - \sum_{e^-} |\hat{\bOmega}\cdot \hat{\mathbf{n}}_e|    \int_e H_i \overline{H}^O_j dA \right]\psi_{gkj}
\end{eqnarray}
Note that the inverse of $\mathbf{T}$ basically sweeps through the mesh for all angles for a given total source $Q^T_{gki}$ (i.e. including inscatter, downscatter, and external sources), and is a linear operator itself.

Then, equation \eref{eq:total_equation_indices} can be rewritten as the following:
\begin{eqnarray}
    T_{gki} - q^I_{gki} = Q_{gki}
\end{eqnarray}
or, equivalently in terms of linear operators as 
\begin{equation}
    (\mathbf{T} - \mathbf{S}\circ \mathbf{M})[\psi_{gkj}] = Q_{gki}    
\end{equation}
The $\phi_{gkj}$ are not necessarily the desired unknowns: the downscatter source depends on the $\phi_{glm}$ instead, and resulting quantities such as tritium breeding or flux only depend on $\phi_{g00}$. Furthermore, $\phi_{glm}$ is much less unknowns that $\psi_{gkj}$; the number of discrete angles is commonly more than a few hundred, while highly anisotropic scattering might only use $L=5$, leading to 36 $lm$ moments. Therefore, we operate first with $T^{-1}$ to obtain 
\begin{eqnarray}
    \psi_{gkj} - \mathbf{T}^{-1}\circ \mathbf{S}\circ \mathbf{M}[\psi_{gkj}] = \mathbf{T}^{-1}[Q_{gki}]
\end{eqnarray}
and thereafter operate with $M$:
\begin{eqnarray}
    \left(\mathbf{M} - \mathbf{M}\circ \mathbf{T}^{-1}\circ \mathbf{S}\circ \mathbf{M}\right)[\psi_{gkj}] = \mathbf{M}\circ \mathbf{T}^{-1}[Q_{gki}]
\end{eqnarray}
to have an equation in terms of the $\phi_{glmj}$ as 
\begin{eqnarray}
    \left(\mathbf{I} - \mathbf{M}\circ \mathbf{T}^{-1}\circ \mathbf{S}\right)[\phi_{glmj}] = b_{glmj}
\end{eqnarray}
where $b_{glmj}:=\mathbf{M}\circ \mathbf{T}^{-1}[Q_{gki}]$, as asserted in equation \ref{eq:a_op_def}.
 \section{Simplex derivation}\label{app:simplices}
 
For simplexes, a mapping to basic element coordinates $\mathbf{q}$ is linear. Then, the Jacobian of this coordinate transformation is a constant and can be taken outside of the integrals.

For an arbitrary function $f$, the volume and edge integrals over the element $M$ and its edge $e$ can be written as an integral over a base element $B$:
\begin{eqnarray}
    \int_M f(\mathbf{r}) dV = \mathcal{J}_v \int_{B} f(\mathbf{r}(\mathbf{q})) d\mathbf{q}\\ 
    \int_e f(\mathbf{r}) dA = \mathcal{J}_e \int_{e_B} f(\mathbf{r}(\mathbf{q})) d\mathbf{q}_e
\end{eqnarray}
where $\mathcal{J}_v$, $\mathcal{J}_e$ are the Jacobians of the transformation to base element (edge) coordinates $\mathbf{q}, \mathbf{q}_e$.
The spatial derivative term can be written as 
\begin{eqnarray}
    \hat{\bOmega} \cdot \nabla H_i = \sum_{\alpha, \beta} \hat{\bOmega}_\beta (\mathbf{J}_v^{-1})_{\alpha \beta} \frac{\partial H_i}{\partial q_\alpha},
\end{eqnarray}
where $\mathbf{J}_v$ is the Jacobian matrix of the transformation to base element coordinates. Now, all terms of equation \eref{eq:total_equation} can be simplified.

For the first term and the last (source) term, note that the basis functions are all defined on the local element coordinates of the base element:
\begin{eqnarray}
    &\int_M \sigma H_i H_j dV = \sigma \mathcal{J}_v \int_B H_i H_j d\mathbf{q} = \sigma \mathcal{J}_v \mathcal{A}_{ij} \\
    \sum_j q_j &\int_M  H_i H_j dV  = \mathcal{J}_v \sum_j  \mathcal{A}_{ij}q_j
\end{eqnarray}
where $\mathcal{A}_{ij}$ does not depend on the element $M$. For the second term:
\begin{eqnarray}
    -\int H_j \hat{\bOmega} \cdot \nabla H_i dV &= \sum_{\alpha, \beta}  \hat{\bOmega}_\beta (\mathbf{J}^{-1}_v)_{\alpha \beta} \int -\frac{\partial H_i}{\partial q_\alpha} H_j dV \nonumber \\
    &= \sum_{\alpha, \beta} \mathcal{J}_v \hat{\bOmega}_\beta (\mathbf{J}^{-1}_v)_{\alpha \beta} \mathcal{B}_{ij\alpha}
\end{eqnarray}
where $\mathcal{B}_{ij\alpha}$ does not depend on the element $M$.

For the third term, since the integration over the edge of the base element depends on the specific edge considered, 
\begin{eqnarray}
    \hat{\bOmega}\cdot \hat{\mathbf{n}}_e \int_e H_i H_j dA = \hat{\bOmega}\cdot \hat{\mathbf{n}}_e  \mathcal{J}_e \mathcal{C}_{ij}^e
\end{eqnarray}
where $\mathcal{C}_{ij}^e$ does not depend on the element $M$ but does on the edge number $e$. 
For the flux propagation term, note that the set of basis functions is the same for all elements. As long as this set of basis functions is invariant under permutations of the nodes (i.e., rotations and reflections), along the common edge, the basis functions in element $O$ can be mapped to the basis functions in element $M$ by transforming the local element coordinates of element $O$. In other words:
\begin{eqnarray}
    \int_{e} H_i \overline{H}_j dA = \mathcal{J}_e \int_{e_B} H_i(\mathbf{q}) H_j(\mathbf{V}(\mathbf{q}))) d\mathbf{q},
\end{eqnarray}
where $\overline{q}$ is the element coordinate in the connected element. Now, along the common edge, this mapping is valid, since both elements contain the edge. It also does not depend on the shape of the total element in real space, only on the connectivity of the common edge. Furthermore, there are only a finite number of ways ($n!$, with $n$ number of vertices) to connect elements of the same type to each other, such that one can compute the matrices $\mathcal{D}_{ij}^{ev}$ for all mappings $\mathbf{V}$ and edges $e$:
\begin{eqnarray}
    \mathcal{D}_{ij}^{ev} = \int_{e_B} H_i(\mathbf{q}) H_j(\mathbf{V}(\mathbf{q})) d\mathbf{q}.
\end{eqnarray}
While setting up the simulation, the specific $e,v$ index can be calculated for all connected boundaries, such that during the sweep the flux propagation term can be quickly computed with a matrix vector product as:
\begin{eqnarray}
      \mathcal{J}_e |\hat{\bOmega}\cdot \hat{\mathbf{n}}_e| \sum_j  \mathcal{D}_{ij}^{ev} \psi^O_j
\end{eqnarray}
In total, the single-element equation is, as before,  
\begin{equation}
    \sum_j G^M_{ij} \cdot \psi_j^M = Q^O_{i} +  Q^M_{i}
\end{equation}
with now
\begin{eqnarray}
    G^M_{ij} &= \sigma \mathcal{J}_v \mathcal{A}_{ij} + \sum_{\alpha,\beta} \mathcal{J}_v \hat{\bOmega}_\beta  (\mathbf{J}_v^{-1})_{\alpha \beta}\mathcal{B}_{ij\alpha}\nonumber \\
    & \ \ \ \ \ \ \ \ \ \ \ \ \ \ + \sum_{e^+}\mathcal{J}_e |\hat{\bOmega}\cdot \hat{\mathbf{n}}_e|  \mathcal{C}_{ij}^e  \\     
    Q^O_i &=\sum_{e^-}\mathcal{J}_e  |\hat{\bOmega}\cdot \hat{\mathbf{n}}_e |\sum_j \mathcal{D}^{ev}_{ij}\psi^O_j  \\
    Q^M_i &=  \mathcal{J}_v\sum_j q_j \mathcal{A}_{ij}  
\end{eqnarray}
where the calligraphic matrices $\{\mathcal{A}_{ij}, \mathcal{B}_{ij\alpha}$, $\mathcal{C}^e_{ij}$, $\mathcal{D}^{ev}_{ij}\}$ do not depend on the element and can be retrieved from a central location.

\section{Anisotropic multigroup scattering data}\label{app:1_scatt_data}
The specific cross sections used for the anisotropic multigroup scattering problem of section \ref{sec:anisotropic} are given in table \ref{tab:app_anisotropic_scatt_data}

\begin{table}[ht]
\caption{\label{tab:app_anisotropic_scatt_data}Macroscopic cross section data for the anisotropic scattering test. Units are in [m$^{-1}$]. Note that the anisotropic part is scaled by a factor $1/(2l+1)$ from Ref. \cite{asaoka1978benchmark} as this is automatically taken into account in both OpenMC and the deterministic models. }
\lineup
\begin{indented}
\item[]\begin{tabular}{@{}l|l|ll|ll}
\br
 & $\sigma^t_g$ [m$^{-1}$]& \centre{2}{$\sigma^0_{g\tilde{g}}$ [m$^{-1}$]} & \centre{2}{$\sigma^0_{g\tilde{g}}$ [m$^{-1}$]}\\
\hline
Water  & 15.2090 & 5.5069 & 0.0      & 1.158567 & 0 \\ 
      & 24.4140 & 6.6227 &11.4080   & 1.1234767 & 2.4733\\\hline
Iron  & 30.3110 & 18.2500 & 0.0      & 4.363 & 0 \\
      & 26.8760 & 6.45294 & 21.3570 & -0.003508 & 2.3457\\\hline
Void  & $1\cdot 10^{-10}$ & 0.0  & 0.0      & 0.0 & 0.0 \\
      & $1\cdot 10^{-10}$ & 0.0  & 0.0 & 0.0 & 0.0\\\hline                                
\end{tabular}
\end{indented}
\end{table}

\end{document}